\documentclass[useAMS,usenatbib]{mn2e}
\usepackage{graphics, graphicx}
\usepackage{amsmath}    
\usepackage{dcolumn}
\newcolumntype{d}{D{.}{.}{-1}}

\newif\ifAMStwofonts

\def\lapp{\ifmmode\stackrel{<}{_{\sim}}\else$\stackrel{<}{_{\sim}}$\fi}
\def\gapp{\ifmmode\stackrel{>}{_{\sim}}\else$\stackrel{>}{_{\sim}}$\fi}

\newcommand{\degrees}[1]{\ensuremath{#1^\circ}}

\def\FirstDFBObservation{2007 April 30 }
\def\LastDFBObservation{2010 June 16}     
\def\TotalNrUniqueDFBObservations{76 }    
\def\TotalUniqueDFBObsrvationHours{7.9 }  

\def\ArchivalDataStart{1998 February 7 }  
\def\ArchivalDataEnd{2007 February 12}    
\def\totalDataSpan{12 }                   
\def\TotalNrArchivalData{256 }            
\def\TotalNrHoursArchivalData{41.9 }      
\def\totalPercentageDouble{0.10\% }       
\def\totalTimeCheckedDouble{49.8 hours }  

\def\TotalInspectedRotations{246,892 }   
\def\PoissonProb{$3\times10^{-7}$ }
\def\TotalInspectedRRATObs{217 }  
\def\TotalInspectedRRATHours{28.3 }  
\def\NrIntegratedPulseObs10word{eight}

\title[The glitch-induced identity changes of PSR J1119--6127]
{The glitch-induced identity changes of PSR J1119--6127}
\author[Weltevrede, Johnston \& Espinoza]
{Patrick Weltevrede$^{1,2}$\thanks{E-mail: Patrick.Weltevrede@manchester.ac.uk}, Simon Johnston$^2$ and Crist\'obal M. Espinoza$^1$\\
$^1{}$Jodrell Bank Centre for Astrophysics, The University of Manchester, Alan Turing Building, Manchester, M13 9PL, United Kingdom.\\
$^{2}$Australia Telescope National Facility, CSIRO, P.O. Box 76, Epping, NSW 1710, Australia.\\
}
\date{}
\pagerange{\pageref{firstpage}--\pageref{lastpage}}
\pubyear{2010}
\begin{document}
\maketitle
\label{firstpage}

\begin{abstract}
Rotation-powered radio pulsars are generally observed to pulse regularly
in the radio band, but this is not the case for so-called rotating
radio transients (RRATs) which emit only sporadic bursts of radio
emission. We demonstrate that the high-magnetic field pulsar
J1119--6127 exhibits three different types of behaviour in the
radio band. Trailing the ``normal'' profile peak there is an
``intermittent'' peak and these components are flanked by two
additional components showing very erratic ``RRAT-like''
emission. Both the intermittent and RRAT-like events are extremely
rare and are preceded by a large amplitude glitch in the spin-down
parameters. The post-glitch relaxation occurs on two different timescales
($\sim$20 and $\sim$210 days) and the post-glitch spin-down rate is
{\it smaller} than the pre-glitch rate. This type of relaxation is also
seen in an earlier, smaller glitch and is very unusual for the pulsar
population as a whole, but is observed in the glitch recovery of a
RRAT. The abnormal emission behaviour in PSR~J1119--6127 was observed
up to three months after the epoch of the large glitch, suggestive of
changes in the magnetospheric conditions during the fast part of the
recovery process. We argue that both the anomalous
recoveries and the emission changes could be related to
reconfigurations of the magnetic field. Apart from the glitches, the
spin-down of PSR J1119--6127 is relatively stable, allowing us to
refine the measurement of the braking index ($n=2.684\pm 0.002$) using
more than \totalDataSpan years of timing data.

The properties of this pulsar are discussed in light of the growing
evidence that RRATs do not form a distinct class of pulsar, but rather
are a combination of different extreme emission types seen in other
neutron stars. Different sub-classes of the RRATs can potentially be
separated by calculating the lower limit on the modulation index of
their emission. Unlike other quantities, this parameter is independent
of observation duration allowing a direct comparison with other
emission phenomenon. We speculate that if the abnormal behaviour in
PSR~J1119--6127 is indeed glitch induced then there might exist a
population of neutron stars which {\it only} become visible in the
radio band for a short duration in the immediate aftermath of glitch
activity. These neutron stars will be visible in the radio band as
sources that only emit some clustered pulses every so many years.

\end{abstract}


\section{Introduction}

PSR J1119--6127 is a young, isolated radio pulsar discovered by
\cite{ckl+00} with the Parkes radio telescope in Australia. Its period
($P=0.41$ s) has a typical value observed for pulsars, but its period
derivative ($\dot{P}=4.0\times10^{-12}$) is among the highest
known. Therefore the inferred surface (dipole) magnetic field strength
is very large ($B_\mathrm{S}=4.1\times10^{13}$ G), making this pulsar
one of the ``high-magnetic field pulsars''. This field strength is
comparable to the so-called ``quantum critical field strength''
of $B_\mathrm{cr}=4.4\times10^{13}$ G\footnote{The surface
magnetic field strength of PSR J1119--6127 at its magnetic pole might
actually be larger than $B_\mathrm{cr}$, because the dipole field there is
twice as strong as the equatorial field $B_\mathrm{S}$.} above which
photon splitting may prevent pair production \citep{bh98}.  Because
this pulsar does not rotate particularly rapidly, the magnetic field
strength at the light cylinder ($B_\mathrm{LC}=5.7\times10^{3}$ G) is
high, but not as extreme as that of other pulsars. From its spin
parameters one can further infer its characteristic age ($\tau_c=1.6$
kyr), which makes its association with supernova remnant G292.2-0.5
likely \citep{cgk+01}.

The high spin-down energy loss rate ($\dot{E}=2.3\times10^{36}$
erg$\,\mathrm{s}^{-1}$) makes this source a promising target for the
Fermi $\gamma$-ray satellite \citep{sgc+08}. Hence it is one of the
energetic pulsars which are currently monitored using the Parkes radio
telescope allowing the detection of possible $\gamma$-ray pulsations
\citep{wjm+10}. During these (ongoing) observations we noted that this
pulsar has striking similarities with a group of neutron stars called
rotating radio transients (RRATs; \citealt{mll+06}) and with the
intermittent pulsars (PSR B1931+24 being the first discovered;
\citealt{klo+06}). This pulsar can therefore be seen, like PSR
B0656+14 \citep{wsr+06}, as an object that links different ``classes''
of neutron stars. We briefly introduce these different types of
objects below.

RRATs are characterized by the sporadic nature of their
emission. Typically, detectable radio emission is observable for less
than one second per day, causing standard periodicity searches to fail
in detecting a periodic signal. Therefore one has to rely on the
greatest common divisor of the time between bursts to derive the
underlying rotational period of the star.
There are about forty reported RRATs in the literature
\citep{mll+06,hrk+08,dcm+09,kle+10,bb10}.
Many RRATs have relatively long periods up to 7 s,
suggesting these sources may be related to the radio-quiet X-ray
populations of neutron stars, such as magnetars \citep{wt06} and
isolated neutron stars \citep{kap08}, which are observed to have
similar periods. The size of the RRAT population is thought to be
several times larger than the radio pulsar population \citep{mll+06},
leading to inconsistencies with the observed supernova rate unless
there is an evolutionary link between the neutron star classes
\citep{kk08}. Different ideas are put forward to explain the sporadic
nature of detectable pulses, such as intermittent particle
precipitation towards the star from a radiation belt similar to those
in planetary magnetospheres \citep{lm07} or re-activation of
the usually inactive vacuum gap due to in falling circumstellar
asteroidal material \citep{cs08}.

\cite{wsr+06} argue that PSR B0656+14 would have been classified as a
RRAT were it not one of the most nearby pulsars. One is
therefore left with the question if PSR B0656+14 should be called a
RRAT? If not, it implies that at least some of the ``RRATs'' are
physically very similar to PSR B0656+14, except that they are more
distant. In any case, it is clear that defining RRATs to be pulsars
that are not detectable via periodicity searches is insufficient and
that one has to define the different neutron star classes based on
physical properties instead (e.g. \citealt{kle+10}).

Based on the similarities between PSR B0656+14 and the RRATs,
\cite{wsr+06} predicted that at least some of the RRATs would show
additional emission which is much weaker than their bright individual
pulses and that when the individual bursts of RRATs are averaged they
will form a ``profile'' much broader than the individual
bursts. Indeed, for instance the RRAT PSR
J0627+16, which was found via its individual pulses, showed weak
emission in a follow-up observation \citep{dcm+09} and the sum of the
individual bursts of the RRAT PSR J1819--1458 produces a triple peaked
pulse profile which is much wider than the individual bursts
\citep{khs+09}. The latter RRAT shows a X-ray spectrum which is
consistent with thermal emission from a cooling neutron star
\citep{rbg+06,gmr+07} and, like PSR B0656+14, it shows X-ray
pulsations \citep{mrg+07}. As will be discussed later in this paper,
the peculiar emission properties of PSRs J1119--6127 and B0656+14
might be linked to the erratic emission seen for other young and
energetic pulsars, such as the Vela pulsar and PSR B1706--44
\citep{jvkb01,jr02}, thereby providing some generalization between the
different phenomenon.

The final class of neutron stars we will discuss in the introduction
are the intermittent pulsars. The archetype is PSR B1931+24
\citep{klo+06}, which is only active for a few days between periods of
roughly a month during which the pulsar is not detectable. These
sudden switches are accompanied by changes in the spin-down rate of
the neutron star rotation, suggestive of significant changes in the
torque generated by magnetospheric currents. It now has become clear
that many more objects show pulse profile changes concurrent with
changes in the spin-down rate \citep{lhk+10} and it has been suggested
that all the so-called timing-noise could be ascribed to this effect.
Indeed the spin parameters of PSR B1931+24 ($P=0.81$ s and
$\dot{P}=8.1\times10^{-15}$) and other physical quantities such as
$B_\mathrm{S}=2.6\times10^{12}$ G, $\dot{E}=5.9\times10^{32}$
erg$\,\mathrm{s}^{-1}$ and $\tau_c=1.6$ Myr are not remarkable,
supporting the idea that the correlated profile and spin-down switches
could be very common in the pulsar population.
It is therefore natural to extend this idea to all
pulsars that switch off (or change their profile) on a regular basis
during so-called ``nulls'' or ``mode changes''. The nulls of some pulsars are very short
(a period or less), while others are only switched on for a few
percent of the time (e.g. \citealt{wmj07}).

In the next sections we give some details about the observations
(Sect. \ref{Sctobservations}) and we will describe a transient profile
change in Sect. \ref{SctIntermittent}. This is followed by the
analysis of the individual pulses (Sect. \ref{SectSinglepulse}),
polarization (Sect. \ref{SectGeometry}) and pulsar timing behaviour
(Sect. \ref{SectTiming}). The paper ends with a discussion of the
properties of PSR J1119--6127 in the light of the different neutron
star classes (Sect. \ref{SectDiscussion}) and the conclusions
(Sect. \ref{SectConclusions}).

\begin{figure*}
\begin{center}
\includegraphics[height=0.45\hsize,angle=270]{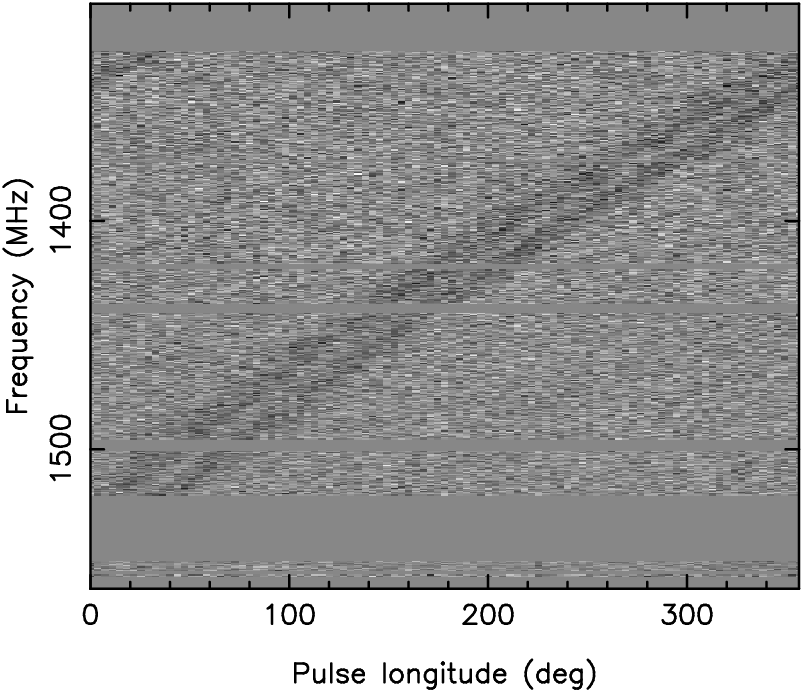}\hspace*{0.05\hsize}
\includegraphics[height=0.45\hsize,angle=270]{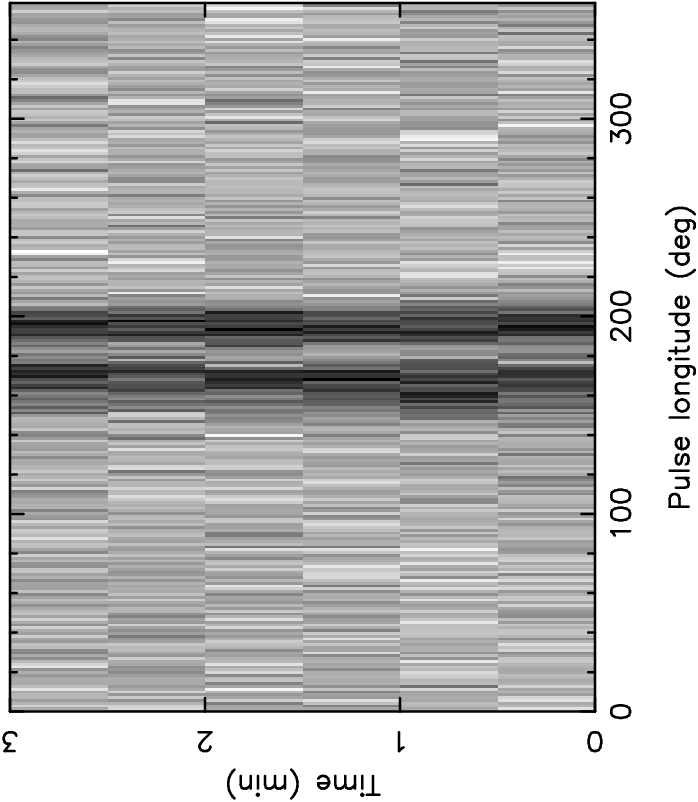}\\
\end{center}
\caption{\label{Figdisp}The only observation known to us showing PSR
J1119--6127 in its double-peaked mode. In both panels the intensity is
shown in grayscale. {\em Left-hand panel:} Both profile components are
identically affected by dispersion (an observing frequency dependent
delay caused by the interstellar medium), resulting in the two
parallel diagonal bands in this panel. {\em Right-hand panel:} The
double-peaked pulse profile was visible during the entire 3 minute
observation. }
\end{figure*}

\section{Observations}
\label{Sctobservations}

All the discussed observations were carried-out using the 64-m Parkes
radio telescope in Australia and the majority of the data were
obtained as part of the aforementioned timing program
\citep{wjm+10}. This program started in April 2007. Each
pulsar is typically observed once per month at 20~cm and twice per
year at 10 and 50~cm, although this pulsar is only detected at 10 and
20~cm. The used receivers were the centre beam of the 20~cm multibeam
receiver (which has a bandwidth of 256~MHz and has a noise equivalent
flux density of $\sim$35 Jy on a cold sky) and the 10/50~cm receiver
(which has at 10~cm a bandwidth of 1024~MHz and has a noise equivalent
flux density of $\sim$49 Jy on a cold sky). The data were on-line
folded at the pulse period by a digital filterbank and were dumped
every 30 seconds on hard disk. For details about the polarization
calibration procedure and the used method to sum all the individual
observations into a high signal-to-noise ``standard'' profile we refer
to \cite{wj08b}. For the majority of the observations the analogue
filterbank was used to record the pulsar signal in parallel with a
time resolution of 250 $\mu$s. Although only total intensity is
recorded it has the advantage that the data are not folded on-line, 
hence allowing the analysis of individual pulses.

In addition we made use of archival Parkes data recorded between
\ArchivalDataStart and \ArchivalDataEnd. For details about these
observations we refer to \cite{ckl+00} in which some of these analogue
filterbank data are published. Together with data from our timing
program up to \LastDFBObservation, this results in a timing history of
more than \totalDataSpan years.

\section{The transient profile component}
\label{SctIntermittent}

The pulse profile of PSR J1119--6127 is well known to be single-peaked
(see e.g. \citealt{ckl+00,jw06}). The observation recorded in 2007
June 17 at 02:35:19 Universal Time therefore immediately triggered our
interest in this object as it shows a clear double-peaked profile (see
Fig. \ref{Figdisp}). Despite years of trying we never observed a
similar event again. Hence, it was important to convince ourselves
that the signal was pulsar related rather than caused by something
instrumental or by terrestrial radio frequency interference (RFI). As
one can see in the left panel of Fig. \ref{Figdisp}, both peaks of the
pulse profile show an identical frequency dependent delay (the gray
horizontal bands are frequency channels affected by RFI and were
excluded from the analysis). This frequency dependent delay is exactly
the same as what is normally observed for this pulsar and can be
perfectly explained in terms of dispersion by the interstellar
medium. This, combined with the fact that the double-peaked structure
is seen during the whole observation after folding at the known pulse
period (see right panel of Fig. \ref{Figdisp}) rules out RFI as the
cause of the additional component. The signal was recorded using two
independent digital filterbank backends\footnote{The used backends are
known as ``DFB1'' and ``DFB2''.} and the recordings look identical as
expected. All observations before and after this particular
observation (of different pulsars) used an identical system set-up and
do not show any sign of a similar effect. Therefore the appearance of
the additional component is clearly pulsar related.

With only one instance of the pulsar showing its double-peaked mode
and no observation of an actual transition it is impossible to
reliably estimate the occurrence rate of this transient
phenomenon. The pulsar was double peaked during the whole 3 minutes of
the observation and the observations 49 days earlier and 35 days later
look normal, implying a total duration of the event of at least
minutes, but shorter than months. In order to maximize the chance of
catching the pulsar again during a similar event we substantially
increased the observing duration (up to half an hour) in subsequent
observing sessions. In total we have observed this pulsar
\TotalNrUniqueDFBObservations times in between \FirstDFBObservation
and \LastDFBObservation, typically twice a day every month with a
total duration of \TotalUniqueDFBObsrvationHours hours. In addition we
inspected archival Parkes data recorded between \ArchivalDataStart and
\ArchivalDataEnd. None of these \TotalNrArchivalData additional
observations (with a total duration of \TotalNrHoursArchivalData
hours) shows a similar event. Therefore the double-peaked mode is
clearly very rare as it is only seen in \totalPercentageDouble of
the inspected data.

\section{Single-pulse analysis: the RRAT-like components}
\label{SectSinglepulse}

\subsection{20-cm observations}

\begin{figure}
\begin{center}
\includegraphics[height=0.85\hsize,angle=270]{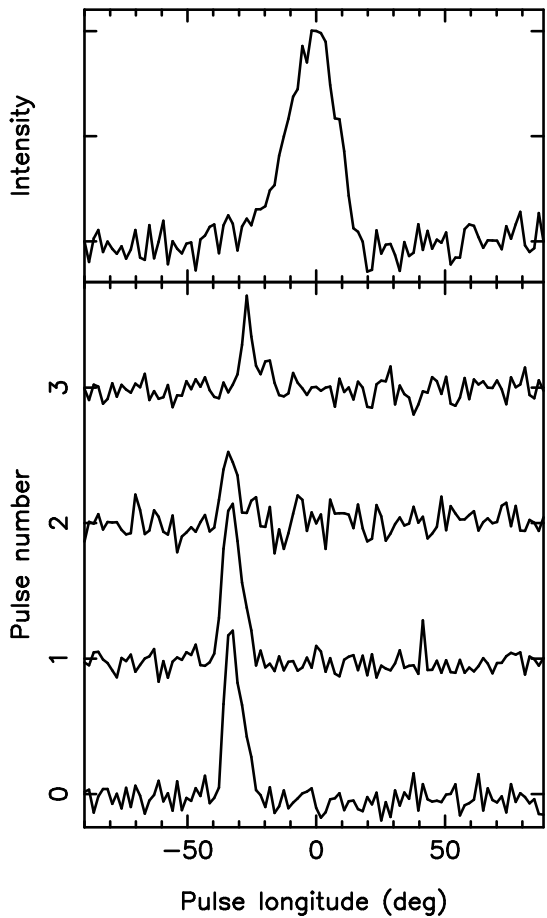}\\
\end{center}
\caption{\label{Figstrongpulses}{\em Top panel:} Averaged pulse
profile of a single 22 minutes long observation at a wavelength of 20 cm. The
time resolution is reduced to 200 bins across the pulse period and the
peak of the profile is put at pulse longitude \degrees{0}. {\em Bottom
panel:} The strong (non-consecutive) individual pulses observed during
the same observation. Notice that at the pulse longitude of the bright
individual pulses there is no integrated power visible, which is
reminiscent of the emission of RRATs.}
\end{figure}

At first glance this pulsar is not a promising source to analyse its
individual pulses because it is quite weak (the average single-pulse
signal-to-noise is less than 1). Nevertheless, careful investigation
by eye of all the observations for which we had single-pulse
recordings available revealed a handful of strong pulses. In total
only four pulses at an observing wavelength of 20 cm were strong
enough to be clearly detectable (see Fig. \ref{Figstrongpulses}). They
all occurred during a 22 minutes long observation recorded at 2007
August 20, starting at 22:11:05 UT.

Strong pulses were visible only in one of the \TotalInspectedRRATObs
observations (with a total duration of \TotalInspectedRRATHours hours)
for which we have single-pulse data (and which were not too badly
affected by RFI). In other words, only during 4 of the
\TotalInspectedRotations inspected neutron star rotations the
intensity of the radio beam was strong enough to be detectable with
the Parkes telescope. Assuming a constant event rate for the
occurrence of strong pulses, Poisson statistics predicts an extremely
low probability of \PoissonProb that four (or more) pulses occur
during a single 3211 pulse period long observation by pure chance.
This strongly suggests that the event rate must have been much
higher during that particular observation.

\begin{figure}
\begin{center}
\includegraphics[height=0.95\hsize,angle=270]{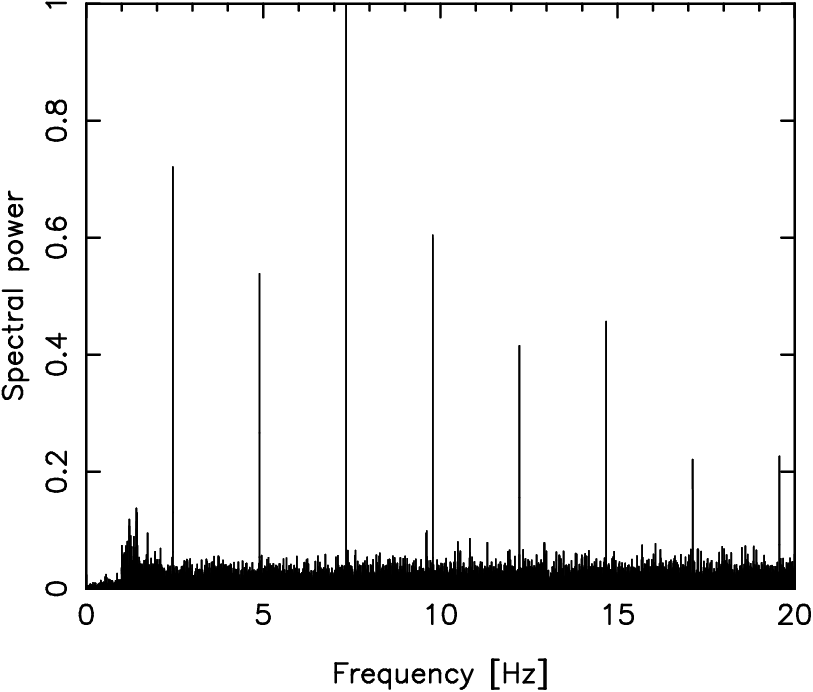}\\
\includegraphics[height=0.95\hsize,angle=270]{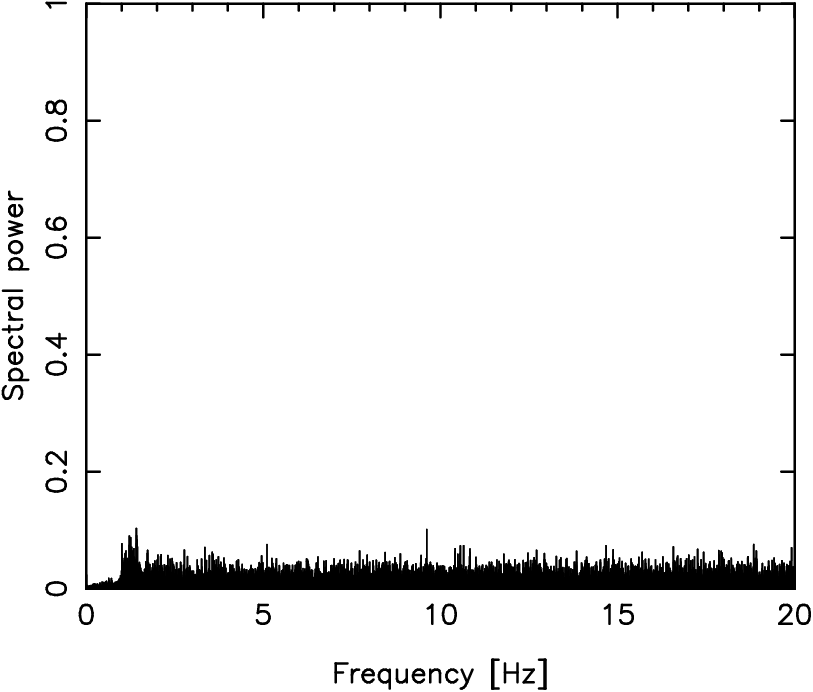}\\
\end{center}
\caption{\label{FigSpectrum}{\em Top panel:} Spectrum of the
time-series of the observation containing the four strong pulses shown
in Fig. \ref{Figstrongpulses}. The periodicity of the pulsar signal clearly
stands out as a spike at 2.45 Hz and spikes at higher frequencies
harmonically related to the fundamental frequency. {\em Bottom panel:}
The spectrum of the same time-series after replacing the signal at
the pulse longitudes corresponding to the normal emission by Gaussian
noise thereby only keeping the strong pulses. None of the visible
spikes correspond to the periodicity of the pulsar. The normalization
of the spectral power is the same in both panels. The decrease in
spectral power below $\sim1$ Hz is due to a software-filter which
removes baseline variations. }
\end{figure}

Looking at Fig. \ref{Figstrongpulses}, one can see that the strong
individual pulses all occur $\sim\degrees{35}$ in pulse longitude
earlier than the peak of the normal profile. Surprisingly, at the
pulse longitudes one can detect individual pulses, the integrated
profile does not show significant emission. This is a defining
characteristic of RRATs, as it implies that this kind of emission
cannot be detected using standard periodicity searches. To prove this
point, we took the observation with the four strong pulses and
replaced the emission within the pulse longitude range corresponding
to the normal emission with Gaussian noise with a root mean square
(RMS) equal to the off-pulse RMS. The spectrum based on this
time-series is shown in the bottom panel of Fig. \ref{FigSpectrum},
while the spectrum including the normal emission is shown in the top
panel. One can clearly see that the fundamental frequency (2.45 Hz)
and the higher harmonics are only visible when the normal emission is
included. Despite the strongest pulses are observed in the leading
component, the underlying periodicity (or the underlying emission
itself) is not detectable at those pulse longitudes. Therefore the
leading component can be qualified to have RRAT-like emission,
although as will be discussed later, there are also important
differences.

\def\BrightnessBrightestSinglePulse20cmRelativeToAveragePulse{23}
\def\FluxBrightestPulse20{18 mJy} \def\PeakFluxBrightestPulse20{0.6
Jy} The strongest of the four detected pulses is
\BrightnessBrightestSinglePulse20cmRelativeToAveragePulse times
brighter than the average pulse of the normal emission during the same
observation. Assuming the normal emission of the pulsar was not
abnormally bright or weak during that particular observation (and
there is no reason to believe so based on the signal-to-noise ratio of
the pulse profile of that observation compared to others), this
corresponds to \FluxBrightestPulse20 averaged out over a whole pulse
period (using the published mean flux density of 0.80 mJy at 1400 MHz
by \citealt{mlc+01}) or a peak flux density of
\PeakFluxBrightestPulse20 using the full single-pulse width of
\degrees{11}.

To investigate how ``normal'' the radio emission in the main-peak
following the RRAT-like component of PSR J1119--6127 is, we attempted
to calculate the modulation index for this pulsar. The modulation
index is independent of the observing system and quantifies the
broadness of an amplitude distribution. It is defined to be
\begin{eqnarray}
\label{EqModIndex1}
m=\frac{\sigma_I}{\left<I\right>}=\left(\frac{1}{N_\mathrm{tot}}\sum_{i=1}^{N_\mathrm{tot}}\left(I_i/\left<I\right>-1\right)^2\right)^{1/2},
\end{eqnarray}
where $\sigma_I$ is the standard deviation of the observed pulse
intensities, $\left<I\right>$ is the average pulse intensity and
$N_\mathrm{tot}$ is the total number of stellar rotations
observed. This quantity can be calculated as function of pulse
longitude by considering the intensities at a specific longitude
rather than the pulse intensities integrated over pulse longitude. We
have chosen to calculate the modulation index in the Fourier domain
instead, which as explained by \cite{es02} allows a correction for
interstellar scintillation.

It turned out that the signal-to-noise ratio in an individual
observation is not high enough to detect a significant modulation
index. To circumvent this problem we have added together all the
individual pulse recording at a wavelength of 20 cm. In order to do
this the average off-pulse intensity of each observation was
subtracted, the data were re-sampled to have an equal number of bins
across the pulse period and the intensities were scaled to make the
off-pulse RMS equal to 1 (thereby assuming that the system temperature
was the same for all observations). The individual observations were
then aligned by cross-correlating their pulse profiles with an
analytic template based on von Mises functions (see
e.g. \citealt{wj08b}). The resulting pulse stack is based on more than 28 hours
of data and shows a modulation index of $m\sim0.5$ in the centre of the
profile, flaring up at the edges. It is likely that the
modulation index is slightly lower in reality because the addition
process can add some artificial variations to the
signal. Nevertheless, the modulation index of PSR J1119--6127 is
entirely consistent to those observed for most pulsars (e.g. \citealt{wes06}).

\subsection{10-cm observations}
\label{Sect10cm}

\begin{figure}
\begin{center}
\includegraphics[height=0.95\hsize,angle=270]{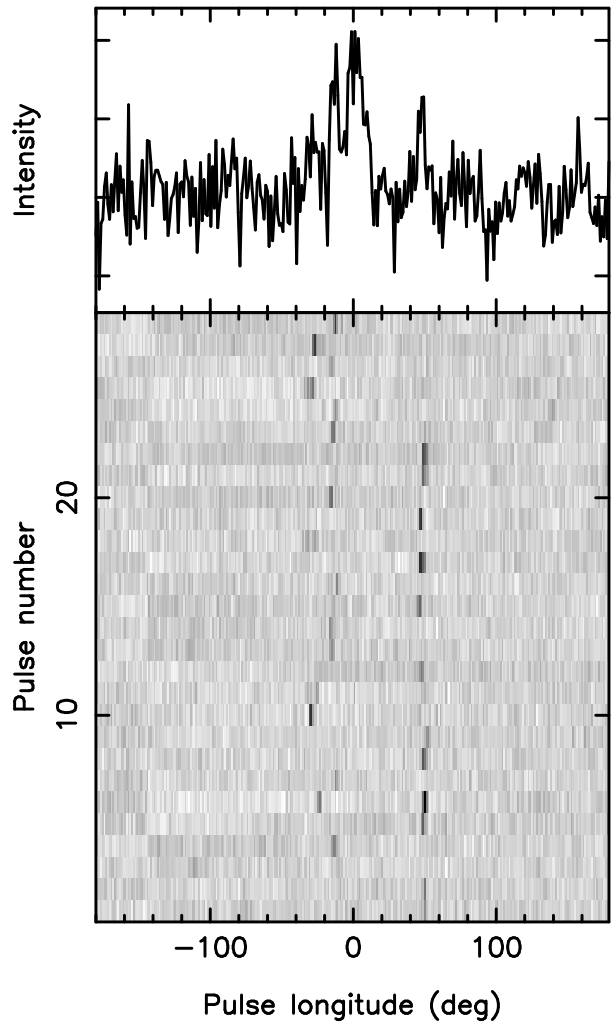}\\
\end{center}
\caption{\label{Figstrongpulses10cm}{\em Top panel:} Pulse profile of
a single 6 minute observation at a wavelength of 10 cm. The time
resolution is reduced to 300 bins across the pulse period. The peak of
the pulse profile is placed at zero longitude. {\em Bottom panel:} The
strong (non-consecutive) individual pulses observed in the same six minute observation are
plotted in grayscale.}
\end{figure}

\begin{figure*}
\begin{center}
\includegraphics[width=0.45\hsize,angle=270]{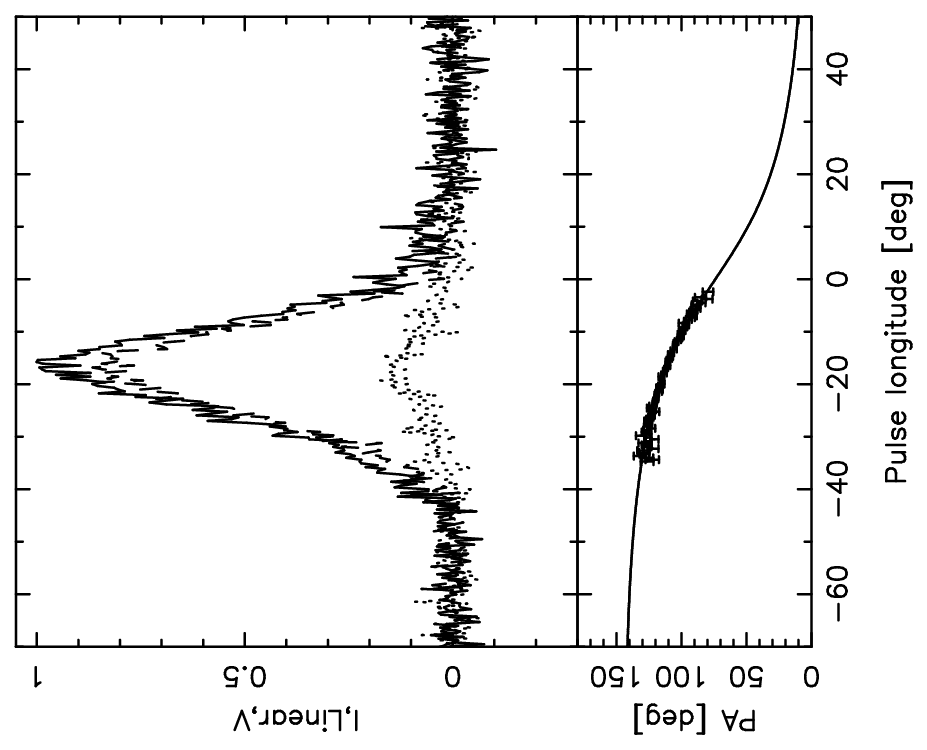}\hspace*{-3.9mm}
\includegraphics[width=0.45\hsize,angle=270]{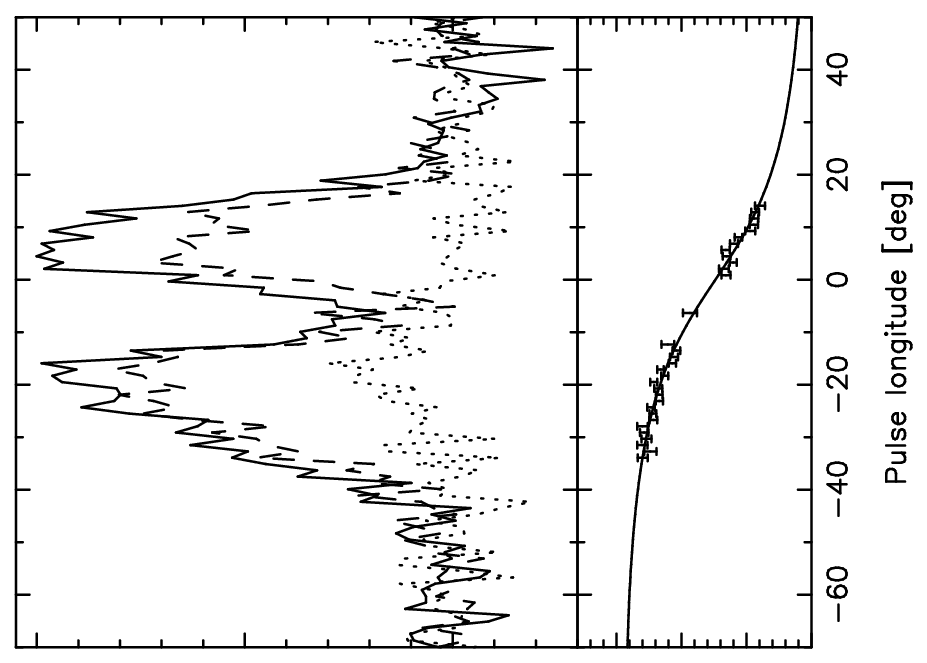}\hspace*{-5.2mm}
\includegraphics[width=0.45\hsize,angle=270]{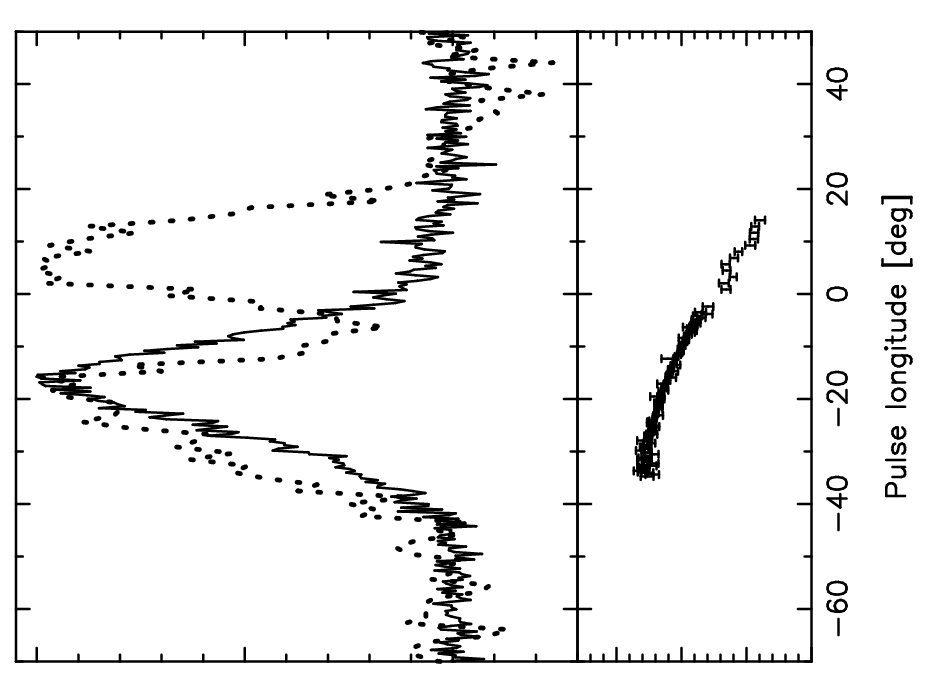}
\end{center}
\caption{\label{paswings}The high signal-to-noise standard profile
showing the pulsar in its single-peaked mode {\em (left panel)} and
the profile in its transient double-peaked mode {\em (middle panel)}.
The top plots show total intensity (solid line), linear polarization
(dashed line) and circular polarization (dotted line). Zero longitude
corresponds to the measured location of the steepest gradient of the
PA curve (shown in the bottom panels). The bottom panel of the middle
panel shows the RVM fit through the PA points. The exact same curve is
shown in the left-hand panel, without any additional fitting
applied. The perfect match, including the absence of any offset
between the data and the model, unambiguously shows that it is the
leading component of the double profile which corresponds to the
single-peaked mode.  {\em Right panel:} The total intensity profiles
of the single-peaked mode (solid line) and the double-peaked mode
(dotted line), as well as their PA points, are overlaid.  The bottom
panel illustrates again the perfect match of the PA-points.  }
\end{figure*}

PSR J1119--6127 has only been observed sporadically at a wavelength of
10~cm during our timing program and at a wavelength of 50~cm the
pulsar is too weak to be detectable for the used integration
times. It must be noted that the combination of a high dispersion
measure, worse system temperature, smaller band width and a much
higher sky temperature might well be preventing us from detecting the
pulsar at 50~cm. Because the individual pulses
were not always recorded at those wavelengths, there are only two
individual pulse recordings available at 10~cm.
One of these recordings, the 6 minute long 2007 July 23 observation,
showed a number of clearly detected bright individual pulses (bottom
panel Fig. \ref{Figstrongpulses10cm}), while a 4 minute long 2009
August 9 observation does not show any sign of similar pulses.

The strong pulses appear to occur at three distinct pulse longitudes
(approximately at $\degrees{-30}$, $\degrees{-10}$ and
$\degrees{50}$). Notice that the pulse profile of this particular
observation shows structure at these pulse longitudes associated with
the strong individual pulses (top panel of Fig.
\ref{Figstrongpulses10cm}). This suggests that the strong pulses
appear before and after the main peak of the observation. The relative
alignment of this profile with respect to the single-peaked and
double-peaked profile observed at a wavelength of 20~cm will be
discussed in the next section in more detail. In total there are
{\NrIntegratedPulseObs10word} observations at a wavelength of 10~cm
for which we have an integrated pulse profile and this observation is
the only one which shows evidence for additional components in the
pulse profile. The fact that the strong pulses do have corresponding
peaks in the pulse profile suggests that at this frequency (or possibly
epoch) the strong pulses are less RRAT-like than the strong pulses
observed at a wavelength of 20~cm.

Independent of whether single pulses occurred in one of the two, or
{\NrIntegratedPulseObs10word}, of the 10-cm observations, given how
rare the detection of individual pulses is at a wavelength of 20~cm
and the fact that at 10~cm more individual pulses are detected
suggests that strong pulses occur more often at higher frequencies.
This is consistent with the fact that there are no individual pulses
detected in two 3 minute observations made at a wavelength of 20~cm
one day earlier and two days later. Only if the event rate is high for
periods of time less than a day and we have been lucky during the
observation at a wavelength of 10~cm the event rates could be the same
at the two frequencies.

\section{Polarization profiles}
\label{SectGeometry}

It is not immediately obvious which of the two components of the
double-peaked profile corresponds to the single-peaked profile and
therefore it is not clear if and how the newly emerged component
relates to the RRAT-like emission. In principle one could use pulsar
timing methods to resolve this issue, but, as will be discussed in
Sect. \ref{SectTiming}, a large amplitude glitch makes this impossible. We
therefore have to resort to other measurements to find out how to time
align the normal single-peaked pulse profile and the double-peaked
profile. We will show in the this section that the position angle (PA)
of the linear polarization can be used to find this relative
alignment. Polarization data are further used to constrain the
viewing geometry and the radio emission height.

\subsection{Alignment of the profile components}
\label{SectAlignment}

Virtually all pulsars with $\dot{E}>2\times10^{35}$ erg\,s$^{-1}$ have
a linear polarization fraction over 50\% (\citealt{wj08b}; see also
e.g. \citealt{hlk98}).
In addition, most of these pulsars show a PA which varies smoothly as
a function of pulse longitude forming a S-shape. This pulsar is no
exception as one can see in the left panel of Fig. \ref{paswings}. The
middle panel shows the double-peaked profile. Notice that the PA
values of the pulsar in its normal mode clearly match those of the
leading peak of the double-peaked profile, suggesting that the
trailing peak is the component which has appeared in the double-peaked
mode. Indeed a cross-correlation between the two profiles shows the
highest degree of correlation for such an alignment. This is caused by
the slight asymmetry seen for the profiles. In the single-peaked mode
the trailing side of the profile is slightly steeper caused by a weak
emission component overlapping with the leading edge of the main
peak (visible at $\degrees{-35}$ pulse longitude). The leading component in the double-peaked mode has also a
shallower leading edge, although in that case there is no evidence for a weak
component being responsible.

From the fact that the PA curve is very similar in the single- and
double-peaked mode one can conclude that the PA provides a much
stronger foundation to base the alignment of the two profiles on than
the total intensity. In order to quantify the relative alignment of
the two profiles we have fitted the data using the rotating vector model
(RVM; \citealt{rc69a}). In this model the PA of the linear
polarization $\psi$ depends on the pulse longitude $\phi$ via
\begin{eqnarray}
\label{EqRVM}
\tan\left(\psi-\psi_0\right)=\frac{\sin\alpha\;\sin\left(\phi-\phi_0\right)}{\sin\zeta\;\cos\alpha-\cos\zeta\;\sin\alpha\;\cos\left(\phi-\phi_0\right)},
\end{eqnarray}
where $\psi_0$ and $\phi_0$ are the PA and pulse longitude at the location
where the $\psi(\phi)$ has its steepest gradient (and an inflection
point). In addition, the shape of the PA curve depends on the emission
geometry of the pulsar, which is described by two angles. Firstly the
angle $\alpha$ between the magnetic axis and the rotation axis and
secondly the angle $\zeta$ between the line-of-sight and the rotation
axis. A related angle is the impact parameter $\beta = \zeta-\alpha$,
which is the angle between the line-of-sight and the magnetic axis at
its closest approach.

According to the RVM, $\phi_0$ corresponds to the pulse longitude at which
the line-of-sight intersects the fiducial plane, the plane containing
the rotation and magnetic axis. There is a longitude-shift one has to take
into account if the emission height $h_\mathrm{em}$ is not negligible
compared to the light cylinder distance, as will be discussed later.
The longitude $\phi_0$ could act as a natural zero point for the pulse
longitude scale if one uses the PA to align different profiles. This
is achieved by determining $\phi_0$ by fitting Eq. \ref{EqRVM} through
the PA points. The profiles are then rotated such that the $\phi_0$
corresponds to \degrees{0} in the plots of Fig. \ref{paswings}.  Note
that the alignment is independent of the actual value of the PA, it
only depends on the pulse longitude corresponding with the steepest
gradient of the RVM curve. This allows a self-consistency check of the
determined alignment by over-plotting the RVM curve obtained from the
fit of the double-peaked profile on the PA points of the single-peaked
profile. One can see in Fig. \ref{paswings} that this RVM curve
matches both PA curves perfectly, something which is also evident from
the right-hand panel which shows the PA points of the two profiles
overplotted. This shows that the measured position of the steepest
gradient of the PA curve is robust, because otherwise there would be a
vertical (or horizontal) offset between the PA points of the two profiles.

\begin{figure}
\begin{center}
\includegraphics[height=0.85\hsize,angle=270]{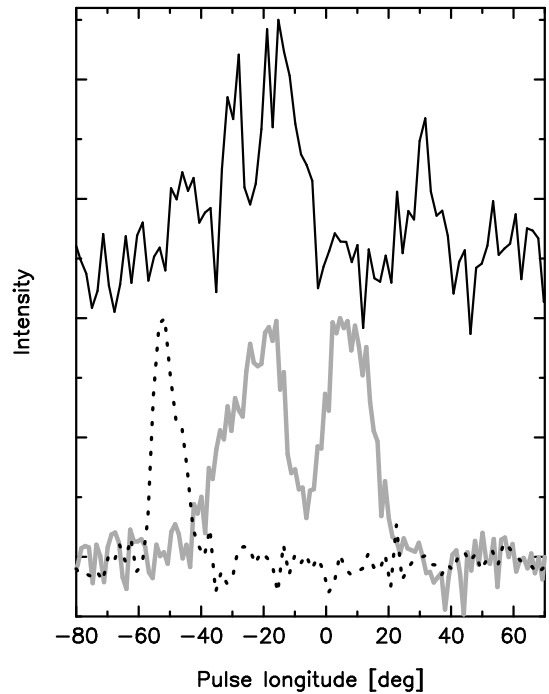}\\
\end{center}
\caption{\label{FigThreeComponents}Plot showing the profiles obtained
by averaging the four strong pulses shown in
Fig. \ref{Figstrongpulses} (dotted line), the double-peaked profile
(thick gray line) and the abnormal profile at 10 cm (solid line, with
a vertical offset from the other two profiles). The pulse longitude
scale is identical to that in Fig. \ref{paswings} and the method for
aligning the 10-cm profile is explained in the text. The peak
intensities of the profiles are normalized.}
\end{figure}

Now that the single-peaked profile is aligned with the double-peaked
profile, the leading RRAT-like component can be placed on the same
longitude scale by aligning the main peaks. This is shown in
Fig. \ref{FigThreeComponents}, which shows both the sum of the four
strong pulses observed at 20~cm (dotted line) and the double-peaked
profile (thick gray line) at the same wavelength. The strong pulses
represent a distinct profile component with a pulse longitude
separation from the main peak roughly equal to that between the main
peak and the transient trailing peak of the double-peaked profile.

The remaining question is how the components seen in the 10-cm
observation discussed in Sect. \ref{Sect10cm} fit in this
picture. The profile is far too weak to find the pulse longitude at
which the gradient of the PA curve is
steepest. Nevertheless, the value of the PA can be used to roughly
align the profile to the PA curve observed at 20 cm. It follows that
the main peak of the 10-cm profile overlaps roughly with the
single-peaked profile observed at 20 cm. In
Fig. \ref{FigThreeComponents} the 10-cm profile is shown together with
the double-peaked profile and the RRAT-like component seen at 20 cm.
This alignment suggests that the bright pulses at the far
leading edge of the 10-cm profile correspond with the RRAT-like
component seen at 20 cm (dotted line). The strong pulses seen at 10 cm
at the far trailing edge of the profile appear to form another profile
component, again with a similar component separation as between the
other components. It should be emphasized that because of the low
signal-to-noise ratio of the 10-cm profile, its alignment with the
other profiles should only be considered as a rough indication.

We therefore believe that although the pulse profile of PSR
J1119--6127 is most of the time single-peaked, there are in fact at least four
components. In addition to the normal emission, the profile has shown
once a transient trailing peak. These two peaks are flanked by two
RRAT-like components which are usually inactive.

\subsection{Emission heights}
\label{SectionEmissionHeights}

\subsubsection{20-cm emission}

The single-peaked profile does not perfectly match the leading
component of the double-peaked profile as shown in the right-hand
panel of Fig. \ref{paswings}. There is no a priori reason why they
should match better, given that the shapes of the profiles are slightly
different to start off with. On the other hand, there are reasons for
the alignment of the profiles not to be perfect. If the emission
height is different in the two states of the pulsar, which potentially
could be the root-cause of the different profile shapes, one does expect a
deviation in the alignment of the profiles. This is because only for a
zero emission height $\phi_0$ corresponds to the pulse longitude at
which the line of sight is passing the fiducial plane. For finite
emission heights co-rotation of the emitting region makes the
inflection point of the PA curve ($\phi_0$) to be delayed with respect to the
pulse profile. The relative shift $\Delta\phi$ between the pulse profile
and the inflection point of the PA curve relates to the emission
height $h_\mathrm{em}$ via (\citealt{bcw91})
\begin{equation}
\label{EqBCW}
h_\mathrm{em} = \frac{1}{4}R_\mathrm{LC}\Delta\phi = \frac{P\,c}{8\pi }\Delta\phi.
\end{equation}
Here $P$ is the spin period of the pulsar, $c$ is the speed of light
and $R_\mathrm{LC}$ is the light cylinder radius (19,500 km for this pulsar).
Therefore, if the emission heights are different in the single- and
double-peaked modes, the pulse profiles will not be perfectly aligned when
using the inflection point of the PA as the reference point.

\begin{table}
 \begin{minipage}{\hsize}
\caption{\label{EmissionHeightTable}This table summarizes the
considered choices for the location of the magnetic axis with respect
to the pulse profiles of PSR J1119--6127. The second and third column
denotes those longitudes in the single- and double-peaked mode
respectively using the longitude shown used in Fig. \ref{paswings}. The next three columns
indicate the implied emission height difference between the two modes
and the emission heights themselves. The last column states which
component of the double-peaked profile is the new component. In case
II the single-peaked profile splits into two components. Both cases I
and IV are argued to be plausible, case IV being our preferred
solution.}
\begin{tabular}[tb]{lccrr@{}lrl}
\hline
\hline
Case\hspace*{-2mm} & $\phi_\mathrm{sgl}$ & $\phi_\mathrm{dbl}$ & \multicolumn{1}{c}{$\Delta h_\mathrm{dbl}$} & \multicolumn{2}{c}{$h_\mathrm{sgl}$} & \multicolumn{1}{c}{$h_\mathrm{dbl}$} & New comp.\\
     &                     &                     & \multicolumn{1}{c}{[km]}                     & \multicolumn{2}{c}{[km]}            & \multicolumn{1}{c}{[km]}\\
\hline
I   & $\degrees{-16}$ & $\degrees{-21}$   &  400  & 1400&\footnote{An emission height of 2200 km was
derived by \cite{jw06} with the steepest gradient lagging the peak of
the total intensity profile by \degrees{26}. We attribute the
discrepancy between their result and ours to the lower quality of
their data and note that their pulse profile and PA curve are fully
consistent with those presented in this paper. An emission height of
2100 km was derived by \cite{wj08b}, however in their table all
emission heights derived from PA curves are overestimated by a factor
2 because of a miscalculation.} &   1800 & Trailing\\
II  & $\degrees{-16}$ & $\degrees{-6}$    & $-900$  & 1400 && 500    & Split    \\
III & $\degrees{-16}$ & $\degrees{+7}$    & $-2000$ & 1400 && $-600$   & Leading  \\
IV  & $\degrees{-6}$  & $\degrees{-6}$    & 0       & 500  && 500    & Trailing \\
\hline
\end{tabular}
\end{minipage}
\end{table}

In order for the centroids to overlap, the normal profile should be
shifted by \degrees{5} to earlier longitudes compared to the
double-peaked profile. This corresponds to an emission height which is
400 km (2\% of the light cylinder radius) lower in the single-peaked
mode (case I in Table \ref{EmissionHeightTable}). Such an emission
height difference would make both the PA-points and the total
intensity profile overlap.  If one allows the emission height
difference between the two modes to be bigger, more extreme offsets
could be considered. For instance, one could envision the fiducial
plane to correspond to the centre of both the single-peaked and the
double-peaked profile, thereby making both profiles symmetric around
the magnetic axis (case II).  Or even more extreme, maybe the single
peak corresponds to the trailing component in the double-peaked mode
(case III). In the final case we consider the magnetic axis
corresponds to the symmetry point of the double-peaked profile without
a differential emission height (case IV).  With the made assumptions
about which pulse longitude of the pulse profile corresponds to the
fiducial plane, Eq. \ref{EqBCW} can also be used to constrain the
emission height itself rather than just height differences. The
resulting emission heights of the single- and double-peaked mode can
be found in Table \ref{EmissionHeightTable}.

Given the fact that the component width is not very different in the
normal mode and the double-peaked mode, one could argue that the
emission height difference between the two modes must be relatively
small compared to the total emission height. For case II (i.e. the
single-peaked profile splits in the double-peaked mode, arguably the
most symmetric scenario) one would expect the overall pulse width to
be roughly 1.7 times larger in the single-peaked mode because of the
curvature of the dipolar magnetic field lines. This effect must be
very precisely counter-balanced by having a smaller fraction of the
polar cap being active during the single-peaked mode in order to have
similar component widths, at which point the model becomes
contrived. This problem becomes even worse in case III, which is
already unlikely given the negative derived emission height in the
double-peaked mode.

In case IV a differential emission height cannot explain the apparent
small offset between the single-peaked profile and the leading
component of the double-peaked profile, as
an essentially zero emission height would be required for the
single-peaked mode. Nevertheless, the striking mirror symmetry of the
double-peaked profile can be seen as strong evidence for the location
of the fiducial plane at that pulse longitude.

Therefore we conclude that, even when differential emission heights
are considered, the most likely scenario is that it is the trailing
peak which has emerged in the double-peaked mode (either case I or IV
in Table \ref{EmissionHeightTable}). Possibly (although not
necessarily) the fiducial plane corresponds to the exact centroid of
both the profile of the single-peaked mode and the leading component
in the double-peaked mode (case I). Otherwise the emission height
might be identical in the two modes with the fiducial plane
corresponding to the centre of the double-peaked profile (case IV).

\subsubsection{10-cm emission}

\begin{figure}
\begin{center}
\includegraphics[height=0.75\hsize,angle=270]{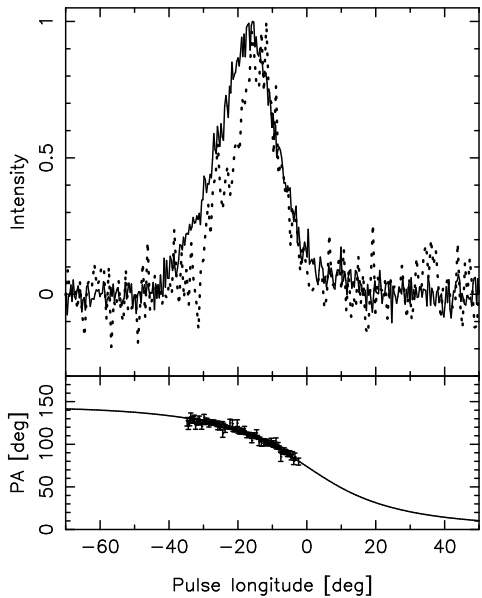}\\
\end{center}
\caption{\label{Fig1020}The normal profile of PSR J1119--6127 at a
wavelength of 20~cm (solid line) and at 10~cm (dotted line). The
steepest gradient of the PA-swing measured at a wavelength of 20~cm
occurs at \degrees{0} longitude. The PA-points of the 10~cm data were
aligned by eye to those at 20~cm and the PA-points at both frequencies
are overlaid in the bottom panel. The alignment of the profiles at
the two frequencies is, within the uncertainties given the relatively
low signal-to-noise ratios, the same as in
Fig. \ref{FigThreeComponents}.}
\end{figure}

Fig. \ref{Fig1020} shows the standard profiles obtained from the
timing program at a wavelength of 20 and 10 cm. The longitude scale
used for the 20-cm profile is the same as in Fig. \ref{paswings} and
the alignment of the 10-cm profile is obtained by matching the PA
curves of the 10- and 20-cm profiles. The centroid of the 10-cm
profile appears to be shifted to later longitudes compared to the 20-cm
profile by about \degrees{3}. Using the same methodology as described
earlier, this suggests that the emission height is $300$ km lower at a
wavelength of 10 cm.
This height difference implies a difference in
the expected half opening angle $\rho$ of the beam  at the two wavelengths when using the
equation
\begin{equation}
\label{EqRho}
\rho=\sqrt{\frac{9\pi h_\mathrm{em}}{2Pc}}.
\end{equation}
In this equation $\rho$ is related to the opening angle of the last
open dipole field lines at the emission height (e.g. \citealt{lk05}),
thereby assuming that the full polar cap is active.

The difference in emission height implies that the emission height at
a wavelength of 20 cm is a factor of 1.2 and 2.0 higher than that of
the 10-cm emission for cases I and IV of Table
\ref{EmissionHeightTable} respectively, which translates in a factor
1.1 and 1.4 difference in $\rho$. The latter factor is similar
to the observed difference in pulse width at the two wavelengths,
therefore supporting scenario IV unless the line of sight is grazing
the edge of the emission cone.
It should be stressed that this is not a very direct argument, as it
involves a chain of assumptions such as the assumption that the same
field lines are producing radio emission at the two
frequencies. Nevertheless, the data are at least consistent with case IV.

\subsection{Viewing geometry}

\begin{figure}
\begin{center}
\includegraphics[height=0.92\hsize,angle=270]{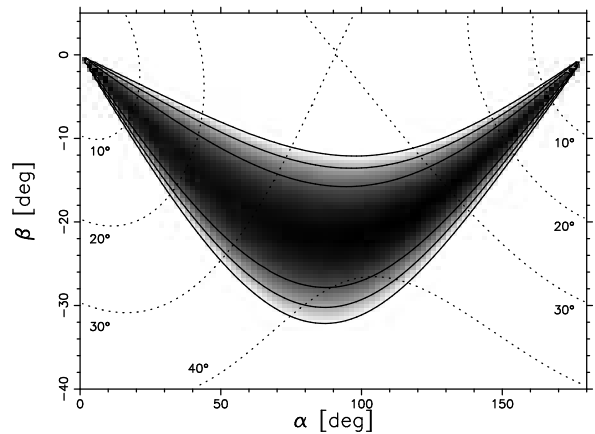}\\
\end{center}
\caption{\label{chi2maps}The $\chi^2$ map resulting from fitting the
RVM model to the polarization data is shown in grayscale. The solid
contours indicate where the $\chi^2$ is two, three and four times
larger than the minimum value. Overlaid are the $\rho$ contours
assuming the double-peaked pulse profile fills the last open field
line region. 
}
\end{figure}

The orientation of the radio beam with respect to the line of sight
and the rotation axis can be quantified with the angles $\alpha$ and
$\zeta$. These angles can be constrained first of all by fitting the
RVM curve (Eq. \ref{EqRVM}). This is visualized in Fig. \ref{chi2maps}
which shows the $\chi^2$ of the RVM fit. As is often the case, the
contours are banana-shaped, caused by a lack in pulse longitude
coverage of the PA points. Therefore, not surprisingly, the best
constraint is obtained by using the PA curve of the double-peaked
profile rather than the single-peaked standard profile despite the
much higher signal-to-noise ratio of the latter.

An additional constraint follows from estimated half opening angle
$\rho$ of the beam (Eq. \ref{EqRho}), which constrains the angles
$\alpha$ and $\zeta$ because $\rho$ is related to the pulse width $W$
via
\begin{equation}
\label{EqCosRho}
\cos\rho = \cos\alpha \cos\zeta+\sin\alpha \sin\zeta \cos\left(W/2\right),
\end{equation}
assuming the radio beam is symmetric about the magnetic axis
(\citealt{ggr84,lk05}).  
The dotted contours in Fig. \ref{chi2maps} indicate the $\rho$ values
as derived using Eq. \ref{EqCosRho}, using the observed pulse width of
the double-peaked profile (which we take to be the full longitude range
over which we observe significant emission).

In case I the radio beam is asymmetric in the double-peaked mode (but
symmetric in the single-peaked mode). Taking the outer edge of the
trailing component of the double-peaked profile to correspond to the
last open field line and using $\rho = \degrees{26}$ (the expected
opening angle for an emission height of 1800 km in the double-peaked
mode) it is derived that $\alpha\sim\degrees{30}-\degrees{40}$.
Note that this implies that the
strong pulses at the trailing edge of the profile observed at 10 cm
are produced outside the polar cap region. If all emission occurs from within
the open field line region the effective pulse width is larger making
$\alpha$ smaller.

For case IV in Table \ref{EmissionHeightTable}, the solution we
preferred in subsection \ref{SectionEmissionHeights}, $\rho =
\degrees{14}$ (Eq. \ref{EqRho}). This suggests
$\alpha=\degrees{20}-\degrees{30}$ (see Fig. \ref{chi2maps}) assuming
the double-peaked profile fills the open field line region. If the
RRAT-like emission comes from open field lines a more aligned geometry
($\alpha\simeq\degrees{17}$) is required to explain the larger overall
pulse width (see Fig. \ref{FigThreeComponents}).

\section{Timing}
\label{SectTiming}

In order to characterise the spin-behaviour of a pulsar one needs to
measure the time-of-arrival (TOA) of the radio pulses as a function of
time. This is done by cross correlating the pulse profiles of the
individual observations with a high signal-to-noise radio standard
profile (shown in the left panel of Fig. \ref{paswings}). Clock
corrections were applied to the TOAs which were then converted to
arrival times at the solar system barycentre using the DE405 model
\citep{ehm06,sta98c} using the TEMPO2 timing package
\citep{hem06}. 

Next, the
barycentric TOAs are compared with a timing model and its parameters
are refined by minimising the so-called timing residuals.  The
refining process has been done using custom software which in essence
works identical to timing packages such as TEMPO2.  The minimisation
procedure makes use of the downhill simplex method and the
uncertainties are derived from the correlation matrix which is
estimated by calculating the Hessian matrix (e.g. \citealt{ptv+92}).

The most basic timing model for the rotation of a pulsar is a 
truncated Taylor series:
\begin{equation}
\label{EqTIMING}
\phi(t)=\phi_0+\nu(t-t_0)+\frac{1}{2}\dot{\nu}(t-t_0)^2+\frac{1}{6}\ddot{\nu}(t-t_0)^3,
\end{equation}
where $\phi(t)$ is the rotational phase as function of time. The
parameters $\phi_0$, $\nu$, $\dot{\nu}$ and $\ddot{\nu}$ are the
rotational phase, spin frequency and its first two time derivatives, all 
defined at epoch $t_0$, usually taken to be during the
time covered by the observations.

PSR~J1119$-$6127 presents a typical timing behaviour of a pulsar
of its age in the sense that like the Crab and other very young
pulsars it exhibits a relatively stable spin-down ($|\dot{\nu}(t)|$)
evolution despite the significant timing noise observed. This allows a
good estimate of its braking index as will be discussed later.
During the time of the observations the rotation of
PSR~J1119$-$6127 has been interrupted by two major glitches exhibiting
a rather particular behaviour. 

A pulsar glitch is a sudden, normally unresolved, increase of the
rotation frequency \citep{rm69}. Sometimes, glitches are observed to
be followed by an exponential-like recovery towards the former
rotational state; phenomenon that has been interpreted as a signature
of the presence of a superfluid in the interior of the star
\citep{bppr69}.  In many cases this recovery is either too small or
the sampling too sparse to quantify its effects.

The two large glitches that we report here occurred in 2004 and 2007;
the first one between MJD~53279 and 53306 (October 1 and 28), and the
second one in 2007 between MJD~ 54220 and 54268 (April 30 and June
17).  All TOAs between MJD~50850 and 55364 were modelled using a
single model that includes these glitches. The phase difference
after a glitch is modelled using an additional phase
$\phi_\mathrm{g}(t)$, which describes the deviation from
Eq. \ref{EqTIMING} caused by the glitch.  The phase after the event is
parametrised by permanent jumps in frequency ($\Delta\nu_p$) and
frequency derivative ($\Delta\dot{\nu}_p$) plus an exponential decay
with time constant $\tau_d$:
\begin{eqnarray}
\label{EqGLITCH}
\nonumber \phi_\mathrm{g}(t)&=&\Delta\phi+\Delta\nu_p(t-t_g)+
                    \frac{1}{2}\Delta\dot{\nu}_p(t-t_g)^2\\
                     &&-\Delta\nu_d\tau_d e^{-(t-t_g)/\tau_d},
\end{eqnarray}
where $t_g$ corresponds to the glitch epoch and $\Delta\nu_d$ is the
step in frequency which is exponentially recovered.  The step in
rotational phase $\Delta\phi$ is not a physical jump in phase, but it
corrects for a $t_g$ which is generally not specified accurate to a
fraction of a rotational period.  The instantaneous frequency jump at
the glitch epoch is given by $\Delta\nu_g=\Delta\nu_p+\Delta\nu_d$ and
the corresponding frequency derivative jump is
$\Delta\dot{\nu}_g=\Delta\dot{\nu}_p+\Delta\dot{\nu}_d$, where
$\Delta\dot{\nu}_d=-\Delta\nu_d/\tau_d$.

\begin{figure}
\begin{center}
\includegraphics[height=0.99\hsize,angle=270]{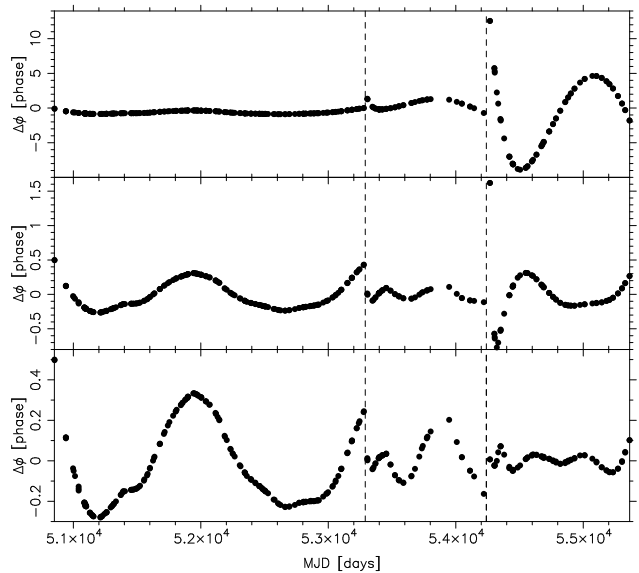}
\end{center}
\caption{\label{timingResiduals}{\em Top panel:} The timing residuals
obtained by only modelling the permanent jumps in rotational frequency
and its time derivative after the two glitches (their epoch is
indicated by the dashed lines). The ordinate is expressed in
turns. {\em Middle panel:} Timing residuals obtained by including a
single exponential glitch recovery after each glitch. {\em Bottom
panel:} Timing residuals obtained by including a glitch recovery on a
second faster timescale after the second glitch.}
\end{figure}

\begin{table}
\caption{\label{TblSPIN}Rotational parameters for PSR~J1119$-$6127.}
\begin{center}
\begin{tabular}{lc}
\hline \hline
Parameter                                                &       Value    \\
\hline
Epoch (MJD)                                             &       52109.85   \\
$\nu$~(Hz)                                              &       2.4512027814(5)    \\
$\dot{\nu}$~($10^{-15}~$Hz\,s$^{-1}$)    &       -24155.07(1)      \\
$\ddot{\nu}$~($10^{-24}~$Hz\,s$^{-2}$)  &       638.9(4)   \\
$\phi_0$                                                  &       37.61(2)   \\
DM~(cm$^{-3}$\,pc)                                 &       713     \\
$n$ & 2.684(2)\\
MJD range                                                &       50850\,--\,55364 \\
\hline
\multicolumn{2}{l}{2004 glitch parameters} \\
\hline
Glitch epoch                                            &       53290   \\
$\Delta\phi$                                           &       4.8(3)     \\
$\Delta\nu_p$~($\mu$\,Hz)                   &       -0.005(13)  \\
$\Delta\dot{\nu}_p$~($10^{-15}~$Hz\,s$^{-1}$)     &       6.4(3)     \\
$\Delta\nu_d$~($\mu$\,Hz)                   &       0.8(1)     \\
$\tau_d$~(days)                                       &       63(8)      \\
\hline
\multicolumn{2}{l}{2007 glitch parameters} \\
\hline
Glitch epoch                                            &       54240   \\
$\Delta\phi$                                           &       22(3)      \\
$\Delta\nu_p$~($\mu$\,Hz)                   &       -1.0(1)    \\
$\Delta\dot{\nu}_p$~($10^{-15}~$Hz\,s$^{-1}$)   &       30(1)     \\
$\Delta\nu_d^{(1)}$~($\mu$\,Hz)            &       5.1(1)     \\
$\tau_d^{(1)}$~(days)                                &       214(7)     \\
$\Delta\nu_d^{(2)}$~($\mu$\,Hz)            &       9(2)       \\
$\tau_d^{(2)}$~(days)                               &       23(3)      \\
\hline
\end{tabular}
\end{center}
\end{table}

\begin{figure*}
\begin{center}
\includegraphics[height=0.99\hsize,angle=270]{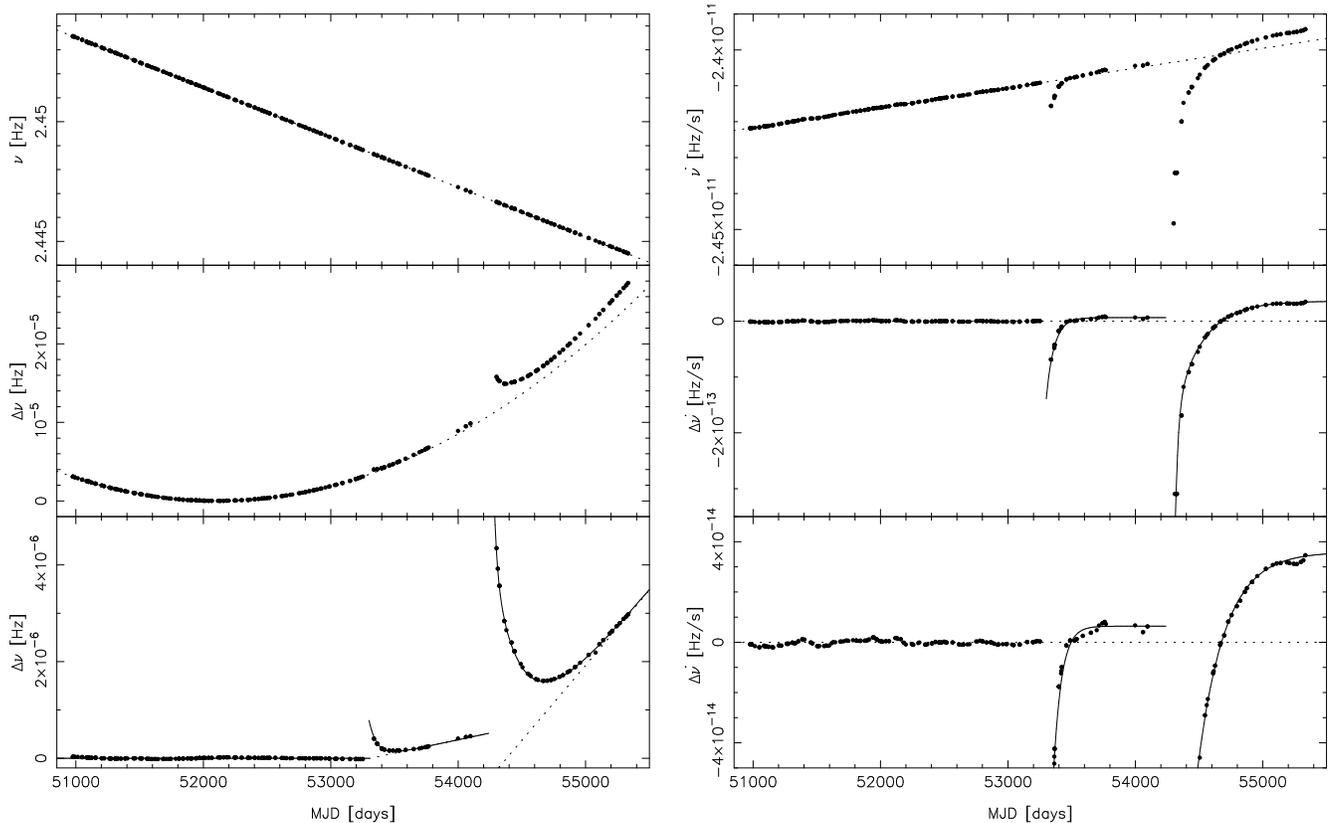}
\end{center}
\caption{\label{timing}{\em Top-left panel:} The measured rotational
frequency of PSR J1119--6127 as a function of time. The frequency is
decreasing, as is expected from spin-down. The prediction according to
the measured $\nu$ and $\dot{\nu}$ in Table \ref{TblSPIN} is shown as
a dotted line. {\em Middle-left panel:} The effect of the constant
slow-down is subtracted from the rotational frequency. Especially the
second glitch can clearly be seen as a deviation from the
parabola-shape. The latter indicates a significant and stable
braking-index and the prediction according to the measured
$\ddot{\nu}$ is shown as a dotted line. {\em Bottom-left panel:} The
difference between the measured spin frequency and the contributions
from $\nu$, $\dot{\nu}$ and $\ddot{\nu}$. The solid and dotted lines
show the prediction according to the glitch model including/excluding
glitch recovery. {\em Top-right panel:} The measured spin-frequency
derivative as a function of time. The dotted line indicates the
$\dot{\nu}$ and $\ddot{\nu}$ contribution. {\em Middle-right panel:}
The difference between the data and the dotted line of the top-right
panel. The solid line indicates the prediction according to the
glitch model including glitch recovery.  {\em Bottom-right panel:}
This plot is identical to the middle-right panel, but slightly zoomed
in. It can clearly be seen that the exponential glitch-recovery is
over relaxing resulting in a positive step in spin-frequency
derivative, i.e. the pulsar spins down slower than what could be
expected from its braking index.}
\end{figure*}

During the minimisation process the epoch of the two glitches were
kept fixed at the values MJD~53290 and 54240, respectively.  The top
panel of Fig. \ref{timingResiduals} shows the timing residuals after
applying the above described model having set $\Delta\nu_d=0$, i.e.
neglecting any possible glitch recovery. It is clear that the
amplitude of the residuals is much larger after both glitches,
indicating that the glitches are poorly modelled. This, in combination
with the observed pattern of the residuals, is suggestive of a
post-glitch transient stage. Accordingly, the residuals are
considerably improved when exponential recoveries are allowed for the
two glitches (middle panel of Fig. \ref{timingResiduals}). Nevertheless,
the residuals immediately after the 2007 glitch show significant
deviations, indicating that the recovery still has not been properly
modelled.  If a second exponential recovery term with a shorter time
constant is included for this glitch, the residuals become much
smaller (bottom panel of Fig. \ref{timingResiduals}). The parameters
of this last fit can be found in Table \ref{TblSPIN}. Because the
recovery process may not yet be complete at the time of writing, the
parameters for the 2007 glitch may change slightly if new data are
added in the near future.

As mentioned earlier, the timing analysis can in principle be used to
identify which component of the double-peaked profile corresponds to
the normal profile. This can be done by calculating a TOA for both
peaks and by determining which corresponding residual follows the
general trend of the residuals closest. However, the double-peaked
profile was detected in the first observation after the 2007 glitch,
hence during the fast part of the recovery process. This makes it
impossible to distinguish between the residuals corresponding to the
two peaks because both residuals can be explained equally well by
slightly adjusting the glitch parameters. The timing solution
presented in Table \ref{TblSPIN} therefore makes use of the alignment discussed
in Sect. \ref{SectAlignment}.

\subsection{Glitch properties}

Fig. \ref{timing} shows the rotational evolution of PSR~J1119$-$6127
over more than 12~years. Its spin parameters as a function of time
were measured using short stretches of data. For each TOA we took all
TOAs within a range of 75 days and minimized the residuals using a
timing model that only included $\phi_0$, $\nu$ and $\dot{\nu}$ as
free parameters and $t_0$ as a fixed parameter.  Fig. \ref{timing}
only includes points which are based on more than three TOAs.

The top left panel of Fig. \ref{timing} shows the frequency against time,
which decreases steadily according to the measured $\dot{\nu}$.
However, the spin-down rate varies slightly with time and the effects
of $\ddot{\nu}$ are only visible when subtracting the effects of
$\dot{\nu}$ first, as shown in the middle-left panel.  The almost
perfect parabola-shape implies a rather constant and stable second
spin-frequency derivative, from which the braking index can be estimated.  In this
panel it is also possible to see the effects of the two large glitches
mentioned above. Both glitches exhibit recoveries towards the
pre-glitch trend, as commonly observed in many young pulsar glitches
\citep{sl96}.
However, it can be observed that especially after the 2007 glitch the 
rotation evolves towards a new state, involving a lower spin-down rate 
$|\dot{\nu}|$ and therefore showing a larger spin-frequency than the 
pulsar would have if it continued evolving as before the glitch (see 
the bottom-left panel).  
This particular behaviour is best appreciated by looking at the evolution
of $\dot{\nu}$, as shown in the top-right panel.  The spin-down rate
evolves almost linearly, as expected for a stable $\ddot{\nu}$, only
interrupted by the two major glitches and their recoveries.  
After each glitch there is a jump in spin-down rate followed by an 
exponential recovery to pre-glitch values.  
However,
as indicated by the positive permanent jump $\Delta\dot{\nu}_p$ 
(Table~\ref{TblSPIN}), after the recovery of the 2007 glitch the 
spin-down rate is smaller than the projected pre-glitch value by 
almost  $30\pm1\times10^{-15}$~Hz\,s$^{-1}$ (see the last two panels at the 
right hand side of Fig. \ref{timing}).
The recovery of the 2004 glitch also presents this behaviour, as
indicated by its positive
$\Delta\dot{\nu}_p=6.4\pm0.3\times10^{-15}$~Hz\,s$^{-1}$.

This post-glitch behaviour is not normal among the rest of the pulsar
population and it has been clearly observed only for the RRAT PSR
J1819--1458 \citep{lmk+09}. 
Normally, if a glitch recovery is detected, $\Delta\dot{\nu}_p\leq0$, 
i.e. the spin-down rate either goes back to values similar to those 
before the glitch or stays at higher rates for a longer time (see Fig.~6 in 
\citealt{lmk+09}). 

The parameter $Q=\Delta\nu_d/\Delta\nu_g$ is used to quantify the
degree of recovery of the spin frequency (e.g. \citealt{sl96}) and it
is observed that pulsars exhibit $Q<1$ for most glitches
(e.g. \citealt{lsg00}).  Because some of the glitch parameters depend
on the glitch epoch, the uncertainty on $Q$ is estimated by obtaining
another 4 solutions (in addition to the one presented in
Table~\ref{TblSPIN}) by varying the glitch epochs in the range defined
by the TOAs just before and after each glitch.  The Q parameters obtained
are both greater than one: $Q=1.01\pm 0.01$ for the 2004 event and
$Q=1.1\pm0.1$ for the 2007 one.  These anomalous values are due to the
negative permanent frequency jumps $\Delta\nu_p$ observed for both
glitches.

A new parameter $Q'=\Delta\dot{\nu}_d/\Delta\dot{\nu}_g$ can be 
introduced to quantify the recovery of the spin-down rate. 
By making use of the relation $\Delta\dot{\nu}_d=-\Delta\nu_d/\tau_d$, 
it can be re-written as
\begin{equation}
 Q' = \frac{\Delta\nu_d}{\Delta\nu_d-\tau_d\,\Delta\dot{\nu}_p}.
\end{equation}
Because $\Delta\dot{\nu}_p$ is usually a negative quantity, 
normal glitches satisfy $Q'<1$.
The $Q'$ values of the 2004 and 2007 glitches are
$1.046\pm 0.004$ and $1.01\pm 0.02$ respectively.
As with $Q$, uncertainties are derived by using the 5 different
timing solutions (see above).
It is worth noticing that if $\Delta\dot{\nu}_p>0$, as is the case for 
these two glitches, then $Q'$ will be always greater than unity. The
amount of overpassing reflects the relative size of the recovered 
jump $\Delta\dot{\nu}_d $ respect to the instantaneous jump 
$\Delta\dot{\nu}_g$. 
Consequently, even though the anomalous recovery is more dramatic
for the 2007 glitch (large $\Delta\dot{\nu}_p$), $Q'$ is larger for the 2004 
glitch.

We note that both $Q$ and $Q'$ are strongly dependant upon the
exponential fit, including the glitch epoch. Changes in the model or
the addition of more post-glitch data could therefore modify their values
considerably.  Nonetheless, the fact they are greater than unity is
significant (despite their relatively large uncertainties), since
it depends purely upon the sign of the permanent jumps
$\Delta\nu_p$ or $\Delta\dot{\nu}_p$ which are measured with much
higher confidence.  In other words, the anomalous behaviour of these two
glitches is already well described by the signs of the permanent
jumps.

We note that \citet{ckl+00} report a small glitch that occurred in
1999 at MJD $\sim$51400. This glitch is much smaller
($\Delta\nu_g=0.011\pm0.001$~$\mu$Hz) than the glitches we report
here.  Although there is clearly a timing deviation at that epoch,
Figs. \ref{timingResiduals} and \ref{timing} do not show convincing
evidence that this glitch is significantly different from other timing
irregularities. Indeed, there was no need to include this glitch in
the long term timing solution (Table \ref{TblSPIN}).

\subsection{Long-term spin evolution}
The long term spin evolution of pulsars can be characterised by the 
so-called braking index $n$, defined by the power law $\dot{\nu}=-k\nu^n$,
where $k$ is a positive constant.
With this definition the braking index can be estimated from the 
observable rotational frequency and its first two derivatives, since
differentiation of the power law gives $n=\nu\ddot{\nu}/\dot{\nu}^2$.
Different braking mechanisms will produce different values of $n$  
\citep{br88}, and it is found that most measured braking indices are less 
than 3,  the canonical value for pure electromagnetic torque 
(e.g. \citealt{lkg+07}).

As it happens with most known very young pulsars, the variation of the 
spin-down rate ($\ddot{\nu}$) is significant and stable, and therefore 
relatively easy to measure.
\citet{ckl+00} measured this variation for PSR J1119$-$6127 and 
reported a braking index $n=2.91\pm 0.05$.
Using the timing solution presented in Table \ref{TblSPIN} we obtain a
braking index $n=2.684\pm 0.002$, the error calculated by propagating
the individual errors on the frequency and its derivatives.  While the
value reported by \citet{ckl+00} was measured using a data span of
about 1.2 yr, this new estimate represents the spin behaviour of the
pulsar for more than 12 years, hence the improvement of the
uncertainty. 
The two large glitches interrupted the spin-down evolution carried on
during the first half of the data (see Fig \ref{timing}), and it is not
obvious how they will affect the long term spin evolution.  If only
data until the first large glitch is used, the result is $n=2.686\pm
0.002$, well contained between the error margins of the value given
above. This similarity makes us confident that the glitches are
modeled in a physically sensible way.

We measure a braking index $n=2.684\pm 0.002$, smaller than the predicted
value for pure dipolar electromagnetic braking, which implies the
presence of extra torques.  The long-term spin evolution of this pulsar
is however unclear, and may not be solely dictated by the measured
braking index.  If glitches like the one reported here are normal, the
effective long-term evolution would have to be described by a larger
braking index. In that case, the quoted value would better represent
the inter-glitch, rather than the long-term spin evolution.

\section{Discussion}
\label{SectDiscussion}

\subsection{The glitch-induced identity changes of PSR J1119--6127}

One of the most intriguing properties of PSR J1119--6127 is that the
radio emission of its distinct profile components can be classified
differently. Its profile is most of the time single peaked and in no
way remarkable. Only once in more than \totalDataSpan years of data a
very different pulse profile was observed with a second transient peak
appearing next to the ``normal'' leading peak. The switching timescale
is therefore more similar to those of the intermittent pulsars than
the shorter timescale mode changes observed in many other pulsars.
The leading component in the transient double-peaked mode is usually
active at radio wavelengths and it is the trailing peak which has
emerged. This identification is made by comparing the PA curve in the
normal and the double-peaked mode. In addition, the circular
polarization and the profile shape show a similar correlation.

Radio emission which can be put into a third category was discovered
by analysing the individual pulses of this source. In a single
observation made at a wavelength of 20 cm four strong pulses leading
both the ``normal'' and ``intermittent'' component were
detected. These pulses are sporadic enough to make the signal
disappear in the noise when adding the emission of all rotations
during the observation together, thereby making this a ``RRAT-like''
component. At a wavelength of 10 cm a second component showing
extremely erratic emission was detected trailing the other profile
components, thereby making these erratic components flank the other
emission.

The ``identity changes'' of PSR J1119--6127 are clearly
glitch-induced. Not only it has been demonstrated that the rate of
occurrence of the RRAT-like pulses at a wavelength of 20~cm must have
been significantly higher around the epoch of their detection, but
more importantly, all abnormal behaviour is observed directly after
the large amplitude glitch. The transient component was detected in
2007 June 17 (once in \totalTimeCheckedDouble of data), which is the
first observation after the glitch. The leading RRAT-like component
was detected at a wavelength of 20 cm in 2007 August 20 (only four
pulses out of the \TotalInspectedRotations inspected stellar rotations in
total). This observation was the first long individual pulse recording
after the glitch at a wavelength of 20~cm (3211 pulses long). Three
other 20~cm observations were made between this observation and the
glitch epoch. However, with a total of only 945 recorded stellar rotations and
the relatively low event rate it is not surprizing the RRAT-like pulses were only
detected during the longer observation.  Finally, the trailing and
leading erratic components were detected in 2007 July 23 at a
wavelength of 10~cm, which is the first 10~cm observation after the
glitch. This makes it likely that erratic RRAT-like pulses were
emitted throughout the time between the glitch epoch and their
detection at 20~cm.

Because of the separation between the observing sessions, the glitch
epoch cannot be accurately determined. As far as we can tell, the
abnormal emission properties of PSR J1119--6127 started directly after
the glitch and continued for $78\pm24$ days.  This corresponds to
$4\pm1$ times the timescale of the fast part of the recovery or
$0.4\pm0.1$ times the timescale of the slow part of the recovery. This
is the first time that a glitch, or the post-glitch recovery process,
is observed to influence the radio emission process of a normal
(non-RRAT) pulsar.

It is currently not possible to know for sure how different the outer
erratic components are compared to the intermittent second peak. Both
phenomenon are observed around the same epoch and appear to be
transient events and could therefore represent different phases of the
same phenomenon. The additional components probably represent regions
on the polar cap which usually fail in producing the positron-electron
pairs required for the emission mechanism to work. One could speculate
that a large-amplitude glitch for some reason changes the
magnetosphere or stellar surface such that these regions switch-on,
either completely or in a flickering manner. Possibly, for all large
glitches the steady emission is followed by a flickering state, before
the physical conditions relax back to the pre-glitch state and the
components switch off again. 
It must be noted that the overall profile symmetry with the more
steady components in the centre and the more flickering components at
the edges of the beam suggests that the different components do have
distinct emission properties rather than we observing the different
components in different phases of the same cycle.

It is not clear why the occurrence of a glitch and the changes in the
radio emission are related to each other. If the glitch is triggered
solely by the conditions inside the star, than how is the energy
involved in the glitch transported to the magnetosphere to induce the
changes in the emission properties? This problem could suggest that
there is an external event (a change in the magnetosphere) that
triggers both the glitch and the changes in the emission
properties. In the next subsection we will argue that the glitches
might be triggered by magnetic stresses on the crust.

\subsection{The glitches of PSR J1119--6127}

\subsubsection{Comparison with other pulsars}

\begin{table*}
\caption{\label{TblQs}Recovery parameters for PSR J1119$-$6127 and
  other pulsars. Uncertainties for the last quoted digit are given in parenthesis. Note that the fact that $Q$ and $Q'$ of PSR J1119$-$6127 are larger than one is significant, despite the large error bars on the values themselves (see text).}
\begin{center}
\begin{tabular}{lldddddl}
\hline \hline
\multicolumn{1}{c}{PSR}  &  
\multicolumn{1}{c}{MJD}  & 
\multicolumn{1}{c}{$\Delta\nu_g/\nu$} & 
\multicolumn{1}{c}{$\Delta\nu_p$} & 
\multicolumn{1}{c}{$\Delta\dot{\nu}_p$} & 
\multicolumn{1}{c}{$Q$}  & 
\multicolumn{1}{c}{$Q'$} & 
\multicolumn{1}{l}{Refs.}  \\
\multicolumn{1}{c}{}     &  
\multicolumn{1}{c}{[days]}  & 
\multicolumn{1}{c}{$(10^{-6})$}  & 
\multicolumn{1}{c}{[$\mu$\,Hz]} & 
\multicolumn{1}{c}{[$10^{-15}$~Hz\,s$^{-1}$]} & 
\multicolumn{1}{c}{} & \multicolumn{1}{c}{} & \multicolumn{1}{c}{}   \\
\hline
J1119$-$6127  &  54240   &  5.4(8)   & -1.0(1)      & 30(1)     &  1.1(1)    & 1.01(2)  &  -     \\
J1819$-$1458  &  53924.8 &  0.699(4) & 0.138(1)     & 0.79(1)   &  0.159(4)   & 1.7(5)    &  1     \\
J1846$-$0258  &  53883   &  4(1)     & -95(1)       & -274(13)  &  9(3)       & 0.973(2)  &  2, 3  \\
B0355+54      &  46497   &  4.368(2) & 27.8773(1)   & -0.5(1)   &  0.00117(4) & 0.14(3)   &  4     \\
B1800--21     &  48245   &  4.07(2)  & 30.07(1)     & -40(1)    &  0.0137(3)  & 0.46(2)   &  4     \\
B2334+61      &  53615   &  20.579(1)& 41.35(3)     & -8.68(2)  &  0.0075(1)  & 0.92(1)   &  5     \\
\hline
\multicolumn{8}{l}{\small {Refs.: 1. \citet{lmk+09}; 2. \citet{lkg10}; 3. \citet{lkgk06}; 4. \citet{sl96}; 5.  \citet{ymw+10} }}
\end{tabular}
\end{center}
\end{table*}

Detected glitch frequency jumps $\Delta\nu_g$ are in the range
0.001--50~$\mu$\,Hz among the pulsar population, implying fractional
changes $\Delta\nu_g/\nu$ between $10^{-11}$ and $10^{-5}$
\citep{sl96,wmp+00}. 
Young pulsars like the Crab and PSR J1119$-$6127 normally exhibit
smaller glitches, with $\Delta\nu/\nu<300\times10^{-9}$.  However, the
2007 glitch, with $\Delta\nu_g/\nu\sim5400\times10^{-9}$, is as big as
the largest glitches detected in the rotation of the Vela pulsar and
many other young pulsars with characteristic ages between 10 and
100~kyr. 

Exponential glitch recoveries have been observed before for other
pulsars and some of them, like PSR J1119--6127, have required more than one
time constant. For instance, more than one timescale was necessary to
model the very large glitch in PSR B2334+61 \citep{ymw+10} and
glitches in the Vela pulsar \citep{mhmk90,accp93}.

A comparison with PSR J1846--0258 is particularly interesting because
its spin parameters ($\tau_c=0.8$~kyr and a magnetic field of
$\sim4-5\times10^{13}$~G) are very similar to that of PSR J1119--6127.
Until now, this rotationally powered radio quiet pulsar was the only
very young pulsars for which a glitch of this magnitude had been
detected.  As observed in PSR J1119$-$6127, the large glitch on that
pulsar presented a very efficient recovery, with $Q=9\pm3$, thanks to
a large negative permanent frequency jump $\Delta\nu_p=-95\pm
1$~$\mu$Hz \citep{lkg10}.  In addition to this anomalous recovery, the
pulsar exhibited a radiative outburst episode in X-rays resembling
magnetar activity, and coincident with the large glitch
\citep{ggg+08,ks08,kh09}.

Glitches are believed to represent changes in the configuration of the
star itself and in general are not associated to external factors.  No
changes on pulse shape or radiative behaviour have been reported for
normal rotation powered pulsars, with the only exception being PSR~J1846$-$0258.
One glitch in the  RRAT PSR J1819$-$1458 also exhibited an anomalous 
spin-down recovery, with 
$\Delta\dot{\nu}_p=0.79\pm 1\times 10^{-15}$~Hz\,s$^{-1}$ and $Q'=1.7\pm 0.5$.
\citet{lmk+09} reported a possible increase in the pulsed
emission rate accompanied by an increase of radio fluxes of the pulses associated with this glitch.

Despite the fact that these three neutron stars have shown anomalous
glitch recoveries, they have somewhat different properties. Table
\ref{TblQs} displays the values $\Delta\nu_p$, $\Delta\dot{\nu}_p$,
$Q$ and $Q'$ for the largest glitches in each of these three
objects.  Additionally, to compare with some other normal glitches, we
also show the same parameters for the very large glitch in the
young PSR B2334+61 \citep{ymw+10}, a large glitch in the Vela-like
pulsar PSR B1800$-$21 and a large glitch in the relatively old PSR
B0355+54.  Values were estimated from published fitted parameters and
references are given in the last column.  While the glitch in PSR
J1119$-$6127 has both $Q$ and $Q'$ greater than unity, the RRAT
exhibits $Q'>1$, but $Q<1$.  On the other hand, the glitch in PSR
J1846$-$0258 presents the opposite situation, with $Q>1$ but $Q'<1$.
The other two normal young pulsars present normal $Q<1$ and $Q'<1$,
owing to normal permanent jumps $\Delta\nu_p>0$ and
$\Delta\dot{\nu}_p<0$.

The emission changes observed in PSR J1119$-$6127 following the 2007
glitch (Sect. \ref{SctIntermittent} and \ref{SectSinglepulse}) and the
anomalous recovery the glitch presented are suggestive of a
single connected event, as observed in PSR J1846$-$0258 and in RRAT
PSR J1819$-$1458.  Both these last objects are high magnetic field
neutrons stars, just as PSR J1119$-$6127.

\subsubsection{Glitch models}

The most simple and intuitive model of a glitch is a starquake
produced by sudden changes of the star shape \citep{bppr69,rud69}.
These changes can be produced regularly as the star slows down and
its shape evolves from an oblate configuration towards a more
spherical one.
However, the high glitch activity of some pulsars (like Vela or 
PSR~B1737$-$30) cannot be accounted for by this model \citep{ml90}.
A second model solved this problem by involving the physics
of the interior of the star.
In this model glitches are produced by a sudden transference
of angular momentum from the inner neutron superfluid to the solid 
star crust.
Any rotating superfluid slows down via migration away of the rotation
axis of quantised vortices; carriers of the superfluid angular momentum.
If the migration is stopped the superfluid retains its angular momentum
and does not slow down.
This model assumes that vortices often pin to different places of the 
inner crust due to interaction with the neutron superfluid associated
with the crustal ion lattice \citep{aaps84a}.
In this way the inner superfluid keeps rotating at at higher rate and
behaves as an angular momentum reservoir.
The collective unpinning of many vortices may occur once stresses
have built up due to the increasing differential rotation between the 
inner superfluid and the crust, producing the sudden spin up of the 
crust we observe as a glitch \citep{ai75,mw09}.

Another model, proposed by \citet{rzc98}, states that vortex migration
would push magnetic flux tubes away, producing stresses in the rigid
crust that may be released through quake-like events.  In addition to
produce possible changes of the moment of inertia of the star, these
reconfigurations could trigger the unpinning of a large number of
vortices, producing a glitch.

Glitches in magnetars present a broad range of properties.  
Among them, the coincidence between radiative outburst and some
glitches is the most relevant, but also the unusual recoveries and
sometimes large post-glitch spin-down increments \citep{dkg08}. 
These characteristics, which are not seen in normal radio pulsars,
have induced the hypothesis that crust rearrangements caused by
magnetic field decay or reconfigurations may trigger these glitches
\citep{dkg07,lkg10}.

\subsubsection{Interpretation of glitch recoveries in terms of the 
pinning--unpinning model of glitches} 
The pinning-unpinning model involves the existence of two main
components: a resistive one, composed of all vortices which are
continuously unpinning and pinning, and a capacitive one, containing
those vortices that unpin only during a glitch \citep{accp96}.  
The resistive component is continuously transferring angular momentum
to the crust, as vortices unpin due to thermal fluctuations or quantum
tunnelling \citep{aaps84,mw09}.
The capacitive component does not participate in the regular spin 
down of the star until a glitch occurs, and the star may contain several
capacitive zones.

During a glitch the crust spins up and the rotational lag with the
superfluid decreases, causing the effects of the resistive component
to decrease \citep{wbl01}.  As a consequence, the spin-down rate
increases due to the action of the always present external torque.
This explains the initial spin-down jump at the glitch epoch and the following
relaxation towards the equilibrium configuration, as vortices from the
resistive component continue to pin and unpin.

Once the recovery is finished one would expect the star to
follow the same spin-down rate as before the glitch.
However, most young pulsars for which a recovery is observed present 
positive permanent jumps in frequency derivative.
For the Crab pulsar, it has been proposed that this change in spin-down is
caused by the creation of vortex trap zones, where many vortices pin and
keep out of the regular slow down \citep{accp96,wbl01}.
The pinning of a large number of vortices decreases the effective moment 
of inertia over which the external torque is acting, hence increasing
the observed spin-down rate. 
In other words, during the glitch many vortices unpin, producing the 
observed spin up, but also a larger number of vortices pin in trap zones.
In this context, $Q$ values less but similar to one are interpreted as a 
capacitive behaviour, where the pulsar is building an angular 
moment reservoir \citep{wbl01}.
\citet{accp96} state that young pulsars like the Crab are still creating 
their capacitive components, which later will produce the larger glitch activity
observed in slightly older pulsars like Vela.
Additionally, they propose that the smaller and less frequent glitches in 
the Crab pulsar are likely to be triggered by starquakes, produced by the 
rapid spin-down of the star.

Following these ideas, and because of their similar age, the glitch
activity in PSR J1119--6127 should present similar characteristics as
the Crab.  However, we have shown that the two last glitches of this
pulsar present anomalous characteristics, specially on their spin-down
rate recoveries.  If the number of vortices that unpinned was
considerably larger than the number of vortices that pinned (during the
creation of capacitive zones), then the effective moment of inertia
after the glitch would be larger than before, and the spin-down rate
smaller.  
This would produce a post-glitch spin-down evolution like the one observed in
Fig. \ref{timing}.  What could cause this different behaviour is not clear,
but the high magnetic field strength is something that PSRs
J1119$-$6127, J1846$-$0258 and RRAT J1819$-$1458 have in common.
Maybe, these glitches have been triggered by magnetic stresses, which
could deform the crust, thereby changing the moment of inertia, or 
release pinned vortices.  Either way, such a glitch would not be part
of the regular glitch activity, and those processes following a normal
glitch may not necessarily occur.  
An unusually large number of vortices may have unpinned in PSR
J1119$-$6127 (considering the low glitch activity of pulsars of its
age), producing a very large spin up.  
But, because the origin of the glitch was different, the unpinning of
vortices overcame any possible vortex pinning, making up a larger
final effective moment of inertia and therefore a smaller spin-down
rate.

\subsection{Comparison with other neutron star populations}

\subsubsection{Comparison with other young pulsars}

The pulse profile of PSR J1119--6127 in its double-peaked mode (see
middle panel of Fig. \ref{paswings}) has striking similarities with a
group of young pulsars with a very distinct pulse shape. These
energetic wide-beam pulsars, such as PSRs J1015--5719, J1105--6107 and
J1420--6048, are characterized by their relatively wide double-peaked
profile with steep inner edges and a high degree of linear
polarization \citep{jw06}. The similarity of this type of profiles
with that of PSR J1119--6127 therefore suggests that the fiducial
plane corresponds to the centre of the double-peaked profile. Having
the magnetic axis in the middle of the double-peaked profile makes it
more natural that the two peaks have the same amplitude when they are
visible.  This scenario (case IV of the discussed geometries in Table
\ref{EmissionHeightTable}) was also preferred after comparing the
pulse profiles at the different observing frequencies.

The main difference between this pulsar and other pulsars of the same
class of profiles might be that the PA curve is not as flat under the
first of the two peaks in the double-peaked mode. The estimated
emission height of $\sim$500 km is somewhat larger than that of the
other pulsars in this group (\citealt{jw06} derived emission heights
of 380, 110 and 175 km for PSRs J1015--5719, J1105--6107 and
J1420--6048 respectively).
The relatively wide observed pulse profile is attributed to
a relatively aligned geometry
($\alpha\simeq\degrees{17}-\degrees{30}$), or possibly smaller if only
part of the open field line region is active.

There are a few young energetic pulsars which exhibit relatively
strong individual pulses, although they are not glitch related. The
strong pulses at the leading edge of the pulse profile of the Vela
pulsar \citep{jvkb01} and at the trailing edge of PSR B1706--44
\citep{jr02} were dubbed ``giant micro-pulses''. These giant
micro-pulses have large peak-flux densities, but their mean flux
densities are not as extreme as the giant pulses observed for instance
for the Crab pulsar (e.g. \citealt{lcu+95}). Unlike normal emission,
the pulse-energy distribution of both giant pulses and giant
micro-pulses are best described by a power law and are therefore
thought to have a somewhat different origin. Unfortunately, not enough
strong pulses were detected for PSR J1119--6127 to quantify their
pulse energy distribution. Nevertheless, there are characteristics the
strong pulses of PSR J1119--6127 do share with the giant micro-pulses:
they all occur at the edges of the pulse profile and are broader than
giant pulses (e.g. \citealt{hkw+03}). Another characteristic that Vela
and PSR B1706--44 have in common is that they are both detected at
$\gamma$-ray energies (e.g. \citealt{aaa+09e}). The $\dot{E}$ of PSR
J1119--6127 suggests that this pulsar might be a $\gamma$-ray source as
well.

\subsubsection{Comparison with the RRATs}

It can be argued that if the line of sight to this pulsar were
different, one would only be able to detect this pulsar via its
individual pulses, hence it would have been classified as a RRAT. This
would suggest that some of the RRATs are very similar in the sense
that they may emit normal radio pulses, but not in the direction of
Earth. There are more analogies one can draw between PSR J1119--6127
and the RRAT population. The unusual post-glitch over-relaxation
observed for PSR J1119--6127 is not normal among the rest of the
pulsar population, but it is similar to that observed for one of the
RRATs \citep{lmk+09}. Also, three of the seven RRATs with a measured
$\dot{P}$ have a $B_\mathrm{S}>10^{13}$ G \citep{mlk+09} and have
therefore field strengths comparable with that of PSR J1119--6127.  In
addition it must be noted that the evolutionary track of
PSR~J1119--6127 in the $P-\dot{P}$ diagram may lead it towards the
RRAT population rather than the bulk of the normal pulsars as is suggested by its braking index.

Having said this, a difference between PSR J1119--6127 and the RRATs
is that in general there is no evidence that the event rate of RRAT
pulses is time-dependent, certainly not at the level as observed for
PSR J1119--6127. One of the RRATs is argued to have an increase in the
RRAT-activity associated with a glitch \citep{lmk+09}, although unlike
PSR J1119--6127 it is always observed to be RRAT-like.  If more
sources like PSR J1119--6127 exist, there might be a population of
neutron stars which emit pulses extremely sporadically after a glitch,
possibly only some clustered pulses every so many years. This kind of
sources, like ``Lorimer burst'' type of events \citep{lbm+07}, would be
hard to detect. It will probably require next-generation telescopes
(such as the LOFAR, ASKAP, MeerKAT or the SKA), which can afford long
dwell times because of their large field-of-view, to discover this
type of object.

PSR J1119--6127 shows erratic emission on both sides of the profile
cone. This might be related to the general observation of the
longitude resolved modulation index being higher at the edges of
profile components (see e.g. \citealt{tmh75,wes06}).  This ``edge
effect'' suggests that the locations of the profile peaks correspond
to field lines at which the emission mechanism is stable. The edge
effect is clearly observed for the Vela pulsar, PSR B1046--58 and
B1706--44 \citep{jvkb01,jr02}, PSR B1133+16 \citep{kkg+03}, and PSR
B0656+14. The latter pulsar shows the strongest pulses in the centre
of the profile (up to 116 times the average pulse intensity), but the
emission on the leading edge can be up to a factor 2000 brighter than
the average emission at those pulse longitudes \citep{wsr+06}. Some
of the RRATs might therefore be pulsars for which the line of sight
only intersects this unstable part of the emission cone.

\subsubsection{Intermittent pulsars}

The double-peaked profile is only observed once in \totalDataSpan
years of data. This timescale is more similar to those of the
intermittent pulsars \citep{lhk+10} than the shorter timescale mode
changes observed in many other pulsars (e.g. \citealt{bac70a,ran86}).
Apart from the similarity in timescale, the most obvious connection
between PSR J1119--6127 and the intermittent pulsars is that changes
in profile shape are linked to the timing properties of the star.
However, the physical mechanism of this link are likely to be
different. In the case of the intermittent pulsars it is believed that
changes in the magnetosphere affect the pulsar rotation (the value of
$\dot{\nu}$), while for PSR J1119--6127 we argue that it is possibly
an external change in the magnetosphere that triggers the glitch.  In
any case, both the intermittent pulsars and PSR J1119--6127 show that
there is an intimate connection between the neutron star interior and
the magnetosphere.

\subsubsection{Comparison with PSR B0656+14}

PSR J1119--6127 shares an important feature with PSR B0656+14. Both
pulsars were discovered using standard periodicity search techniques,
but subsequently found to exhibit RRAT-like emission. Both pulsars
therefore provide a link between the ``normal'' pulsars and the
RRAT-like pulsars. Pulsed X-rays have been observed for both PSR
J1119--6127 \citep{gkc+05} and PSR B0656+14 \citep{chmm89,dcm+05}.  In
the radio band they both have a more or less triangular-shaped pulse
profile which covers only a part of the open field
line region (in the case of PSR B0656+14 there is a trailing shoulder,
while PSR J1119--6127 has the intermittent trailing peak). They both
have a high degree of linear polarization at 1.4 GHz and it will be
interesting to see if PSR J1119--6127 becomes depolarized at higher
frequencies, like PSR B0656+14 \citep{jkw06}.

There are also some important differences between PSRs J1119--6127 and
B0656+14. PSR B0656+14 is a middle-aged pulsar located near the bulk
of pulsars in the $P$-$\dot{P}$ diagram, while PSR J1119--6127 is very
young and has an extremely strong magnetic field. The RRAT-like
emission of PSR J1119--6127 is found in an separate profile component
(more like giant micro-pulses), while the strongest pulses of PSR
B0656+14 are found at the same pulse longitude as the normal
emission. Therefore the RRAT-like emission of PSR B0656+14 is
interpreted as being the tail of an extended pulse-energy distribution
\citep{wws+06}, while in the case of PSR J1119--6127 and the giant
micro-pulses the RRAT-like emission is more distinct from the normal
emission.  Also, the emission of PSR B0656+14 was found to be more
extreme at low observing frequencies, while for PSR J1119--6127 strong
pulses possibly occur more often at higher observing
frequencies. Therefore the physics of the strong pulses observed for PSR
B0656+14 might be more related to those observed for older pulsars at
very low frequencies (e.g. \citealt{ek05}).  In addition to these
differences the strong pulses of PSR B0656+14 are not related to
glitches.

\subsubsection{Comparison with magnetars}

Because of the high magnetic field strength of PSR J1119--6127 there
is an obvious direct link with the magnetars.  Links between magnetars
and rotation-powered pulsars have been suggested before as for
instance PSR J1846--0258, a young (radio-quiet) pulsar with a similar
magnetic field strength to that of PSR J1119--6127, has shown
magnetar-like X-ray bursts \citep{ggg+08,ks08}. In addition, an
over-recovery ($Q=8.7\pm2.5$; \citealt{lkg10}) has been claimed for
that source, as well as for magnetar AXP 4U 0142+61 ($Q=1.07\pm0.02$;
\citealt{gdk09}).

Three of the magnetars are detected as radio pulsars
\citep{crh+06,crhr07,lbb+10} and they show characteristics similar to
PSRs J1119--6127 and B0656+14.  In particular all three have a high
degree of linear polarization \citep{crj+07,ksj+07,crj+08} and
individual pulses from PSR J1810--197 \citep{ssw+09} and J1623--4950
\citep{lbb+10} show spiky emission similar to that seen in PSR
B0656+14 \citep{wws+06}. Curiously, evidence seems to suggest that the
magnetars are close to being aligned rotators \citep{crj+07,crj+08}
and this might also be the case for PSR~1119--6127 and possibly PSR
B0656+14 \citep{ew01}.  Could it be that the aligned geometry is
responsible for some of these attributes?  We note also that the
$\gamma$-ray profile of PSR~B0656+14 is peculiar \citep{waa+10} and we
await the detection of PSR~J1119--6127 with the Fermi satellite with interest.

\subsection{What are RRATs?}

The inconsistency between the observed supernova rate and the size of
the RRAT population suggests there are links between the different
neutron star classes \citep{kk08}. Indeed, it has been argued that one
of the RRATs, PSR J1819--1458, might be evolved from the magnetars
based on the unusual $\dot{P}$ change after its glitch \citep{lmk+09}.
Because of the movement of the Crab pulsar in the $P-\dot{P}$ diagram
the young pulsars and the magnetars appear to be linked as
well \citep{lyn04}. 
The view that the RRATs are an evolutionary stage rather than a
separate class of neutron stars is further strengthened by the discovery of
``missing links'': pulsars which can be classified both as pulsars or
as RRATs. Some RRATs will turn out to be pulsars with extended pulse
energy distributions (like that of PSR B0656+14; \citealt{wsr+06}),
others might have a ``normal'' profile component (like PSR
J1119--6127) which happen to be missed by our line of sight, others
might be pulsars with extreme null lengths (see e.g.  \citealt{bb10})
and some might be magnetars with more extreme spiky radio emission
than XTE J1810--197 \citep{ssw+09}.

With the above in mind, a valid question to ask oneself is what you
are left with when you remove the pulsars that fall in the above
categories from the RRAT population. Will there be any ``true'' RRATs
left? Or are RRATs simply the combination of the extreme ends of the
other classes of neutron stars?  
Because RRATs do not form a separate island in the $P-\dot{P}$ diagram
and because it proves to be difficult to define the RRATs based on
their physical properties (see e.g. \citealt{kle+10}), it is tempting
to apply Occam's razor and assume that the RRATs do not form a
distinct population of pulsars.

A problem with the term RRAT is that it is not clearly defined. The
definition that RRATs are pulsars which are not detectable via
periodicity searches is insufficient because (as illustrated by PSR
B0656+14) it depends on the distance of the source (and the
integration time of the observation and the sensitivity of the
telescope).
By accepting that RRATs are not a separate class of pulsars the
question of how to define a RRAT is irrelevant. More useful would be
to define how erratic (hence ``RRAT-like'') the emission of a pulsar
is. Such a definition would allow a better investigation of how this
type of emission is correlated with other neutron star parameters.

The degree of RRAT-like behaviour can be quantified by intermittency
ratio \citep{mc03,dcm+09,kle+10}, which compares the
signal-to-noise ratio of the detection of the pulsar in a single-pulse
search with that of a periodicity search, or the R-parameter
\citep{jvkb01}, which is a normalized measure of the brightest
observed pulse. Both measures have the problem that it depends on both
the sensitivity of the telescope and the duration of the
observation. A straight-forward way to define the degree of RRAT-like
behaviour would be to use the modulation index
(Eq. \ref{EqModIndex1}), which quantifies the broadness of an
amplitude distribution and is observing-system independent. Note that
$m$ is usually calculated as a function of pulse longitude. This makes
a big difference for especially the magnetar emission, which shows
many bursts per stellar rotation.

For RRATs
$\left<I\right>$ is below the detection threshold and therefore the
modulation index cannot be calculated (the same problem applies for
the intermittency ratio or the R-parameter). Nevertheless one can
determine a lower limit on the modulation index by taking zero
intensity for the undetected pulses and replacing $\left<I\right>$ by
its maximum allowed value, which is the detection threshold for the
average pulse intensity $\left<I\right>_\mathrm{tresh}$. The resulting
limit is
\begin{eqnarray}
\label{EqModIndex2}
m\geq\left(\frac{1}{N_\mathrm{tot}}\sum_{i=1}^{N_\mathrm{det}}\left(I_i/\left<I\right>_\mathrm{tresh}-1\right)^2\right)^{1/2},
\end{eqnarray}
where $N_\mathrm{det}$ is the number of detected pulses.

\begin{table}
\begin{center}
 \begin{minipage}{\hsize}
\caption{\label{ModIndexTable}Table with some typical values of the modulation index $m$ measured for different types of emission.}
\begin{tabular}[tb]{llcc}
\hline
\hline
Source & Type of emission & $m$ & Ref.\\
\hline
Typical pulsar            & Normal  & 0.5 & \footnote{Typical minimum value of the longitude resolved modulation index at a wavelength around 21 cm \citep{wes06} and $m$ is typically observed to flare up at the profile edges.}\\
Crab pulsar & Giant pulses & $5-8$ & \footnote{Longitude resolved modulation index at the position of the giant pulses at 1400 MHz \citep{wes06}. The actual value might be higher if the giant pulses were not entirely resolved.}\\
PSR B0656+14 & RRAT-like  & $\sim5$ & \footnote{Measured by \cite{wws+06} at a frequency of 327 MHz in the
centre of the profile (where the brightest burst occurred). }\\
PSR J1119--6127 & Normal  & $\sim0.5$ & \footnote{This paper. The modulation index flares up at the profile edges.}\\
PSR J1119--6127 & RRAT-like  & $\geq0.6$ & \footnote{Modulation index of the RRAT-like component based on the single 20-cm observation with the four individual pulses.}\\
XTE J1810--197 & Magnetar emission & $1-10$ & \footnote{The longitude resolved modulation index is variable in time, but in the range of $m=1-4$
at 1.4 GHz or $m=1-10$ at 8.35~GHz \citep{ssw+09}.}\\
PSR J1819--1458 & RRAT & $\geq6.6$ & \footnote{This is the RRAT with the best
determined pulse energy distribution. The estimated limit is based on this distribution (Supplementary Figure 2
of \citealt{mll+06}) in combination with the quoted event rate of 229
detected pulses in 13 hours of data and the statement that average
peak flux density must be less than 0.05\% that of the strongest
detected burst. }\\
\hline
\end{tabular}
\end{minipage}
\end{center}
\end{table}

The modulation index for selected sources can be found in Table
\ref{ModIndexTable}. One can see that the modulation index of the
normal emission of PSR J1119--6127 is typical for that of radio
pulsars. Unfortunately, the lower limit of the modulation index of the
RRAT-like emission of PSR J1119--6127 is not much higher than that of
the normal emission, and therefore it is consistent with all types of
emission. To improve this limit one needs to use an instrument which
is much more sensitive than Parkes, such as the SKA. On the other hand,
one can see that the modulation index of RRAT J1819--1458 must be much
higher than that of normal pulsar emission and is possibly more
similar to that found for magnetar emission or that of giant
pulses. Measurements of $m$ for more RRATs and other sources with
extreme radio emission has the potential to differentiate between
different types of radio emission.

\section{Conclusions}
\label{SectConclusions}

We found that the young high-magnetic field pulsar J1119--6127
exhibits three types of pulsar behaviour. It has an ``intermittent''
profile peak trailing the peak associated with ``normal'' emission and
these profile components are flanked by two much more erratic
components, which can be argued to be ``RRAT-like''. Both the erratic
components and the intermittent component are observed around the same
epoch and appear to be transient events. It can therefore not be ruled
out that they represent different phases of the on- and off-switching
process, possibly related to regions on the polar cap which usually
fail in producing the positron-electron pairs required for the
emission mechanism to work. Nevertheless the overall profile symmetry
suggests that the more steady components are located in the centre of
the beam and the more flickering components at the edges. This is
consistent with the ``edge effect'', which states that in general the
emission is more erratic at the edges of the profile components. Some
of the RRATs might therefore be pulsars for which the line-of-sight
misses the part of the beam with more steady emission.

Both the intermittent and RRAT-like events are very rare and are
preceded by a large amplitude glitch that occurred directly before
these phenomenon were observed. This is the first time that a
glitch, or the post-glitch recovery process, is observed to influence
the radio emission process of a normal (non-RRAT) pulsar.  The glitch was
followed by a post-glitch behaviour which is very unusual for the
pulsar population as a whole, but it is similar to that observed for
one of the RRATs (PSR J1819--1458).  After the glitch, the spin-down
rate relaxed to a smaller value than the projected pre-glitch value.
We interpret this as an increase of the effective moment of inertia
produced by an excessive number of unpinning vortices. By considering
this anomaly and the emission changes observed, we believe that the
glitch may have been triggered by magnetic reconfigurations.

Although we measure a braking index $n=2.684\pm 0.002$, glitches like
the one reported here make the effective braking index
larger. Nevertheless, the evolutionary track of PSR~J1119--6127 in the
$P-\dot{P}$ diagram may lead it towards the RRAT population rather
than the bulk of the normal pulsars. The occurrence-rate of RRAT-like
pulses must have been higher around the glitch epoch (as is claimed
for J1819--1458), suggesting a link with the glitch-recovery
process. More precise, the RRAT-like pulses were observed during the
initial fast recovery (on a timescale of $\sim21$ days) rather than
the later slow recovery (on a timescale of $\sim208$ days).  This
allows the possibility that RRAT-like activity is in general low for
young, high-magnetic field pulsars (except after a glitch) and
increases when pulsars age.

Like PSR B0656+14, which would be classified as a RRAT were it more
distant, PSR J1119--6127 provides a link between the known neutron
star population and the RRATs. Additional links could be other young
pulsars which show ``giant micro-pulses'', the magnetar population
which show PSR B0656+14-like spiky emission and pulsars with extreme
long null lengths. It appears that RRATs represents a corner in the
$P-\dot{P}$ diagram in which the pulsars have the most erratic
emission.
We argue that the modulation index is a good measure of how RRAT-like
the emission is. The modulation quantifies the broadness of the pulse
energy distribution and is (unlike for instance the ``intermittency
ratio'' or the ``R-parameter'') independent of the sensitivity of the
telescope or observation duration. For RRATs (which are not detected
in a periodicity search) one can only derive a lower limit. This limit
is estimated to be $m\geq6.6$ for the RRAT PSR J1819--1458, which
although high for pulsar standards is not necessarily higher than that
observed for magnetars.
Additional observations could place stronger
limits on how high the modulation index of RRAT emission is and
can potentially be used to distinguish between the
different sub-classes of the RRATs.

Not only are the strong individual pulses of PSR J1119--6127
reminiscent of those of other young pulsars, but so is the shape of
its pulse profile in its double-peaked mode. The similarity with other
``energetic wide-beam pulsar'' suggests that the magnetic axis is
located in between the two components. The derived emission height for
that scenario is $\sim$500 km, somewhat higher than that derived for
other young pulsars.  thereby suggesting a relatively aligned geometry
($\alpha\sim\degrees{17}-\degrees{30}$), or slightly more aligned if
only part of the open field line region is ever active.

\section*{Acknowledgments}
The authors want to thank Ben Stappers and Evan Keane for the useful
discussions. The Australia Telescope is funded by the Commonwealth of
Australia for operation as a National Facility managed by the CSIRO.

\bibliographystyle{mn2e}

\begin{thebibliography}{}

\bibitem[\protect\citeauthoryear{{Abdo}, {etal1} \& {etal2}}{{Abdo}
  et~al.}{2010}]{aaa+09e}
{Abdo} A.~A., et al. 2010, Astrophys. J. Supp. Series, 187, 460

\bibitem[\protect\citeauthoryear{Alpar, Anderson, Pines \& Shaham}{Alpar
  et~al.}{1984}]{aaps84a}
Alpar M.~A.,  Anderson P.~W.,  Pines D.,    Shaham J.,  1984, ApJ, 276, 325

\bibitem[\protect\citeauthoryear{Alpar, Chau, Cheng \& Pines}{Alpar
  et~al.}{1993}]{accp93}
Alpar M.~A.,  Chau H.~F.,  Cheng K.~S.,    Pines D.,  1993, ApJ, 409, 345

\bibitem[\protect\citeauthoryear{Alpar, Chau, Cheng \& Pines}{Alpar
  et~al.}{1996}]{accp96}
Alpar M.~A.,  Chau H.~F.,  Cheng K.~S.,    Pines D.,  1996, ApJ, 459, 706

\bibitem[\protect\citeauthoryear{{Alpar}, {Pines}, {Anderson} \&
  {Shaham}}{{Alpar} et~al.}{1984}]{aaps84}
{Alpar} M.~A.,  {Pines} D.,  {Anderson} P.~W.,    {Shaham} J.,  1984, ApJ, 276,
  325

\bibitem[\protect\citeauthoryear{Anderson \& Itoh}{Anderson \&
  Itoh}{1975}]{ai75}
Anderson P.~W.,  Itoh N.,  1975, Nature, 256, 25

\bibitem[\protect\citeauthoryear{Backer}{Backer}{1970}]{bac70a}
Backer D.~C.,  1970, Nature, 228, 1297

\bibitem[\protect\citeauthoryear{Baring \& Harding}{Baring \&
  Harding}{1998}]{bh98}
Baring M.~G.,  Harding A.~K.,  1998, ApJ, 507, L55

\bibitem[\protect\citeauthoryear{Baym, Pethick, Pines \& Ruderman}{Baym
  et~al.}{1969}]{bppr69}
Baym G.,  Pethick C.,  Pines D.,    Ruderman M.,  1969, Nature, 224, 872

\bibitem[\protect\citeauthoryear{Blandford \& Romani}{Blandford \&
  Romani}{1988}]{br88}
Blandford R.~D.,  Romani R.~W.,  1988, MNRAS, 234, 57P

\bibitem[\protect\citeauthoryear{Blaskiewicz, Cordes \& Wasserman}{Blaskiewicz
  et~al.}{1991}]{bcw91}
Blaskiewicz M.,  Cordes J.~M.,    Wasserman I.,  1991, ApJ, 370, 643

\bibitem[\protect\citeauthoryear{{Burke-Spolaor} \& {Bailes}}{{Burke-Spolaor}
  \& {Bailes}}{2010}]{bb10}
{Burke-Spolaor} S.,  {Bailes} M.,  2010, MNRAS, 402, 855

\bibitem[\protect\citeauthoryear{Camilo, Kaspi, Lyne, Manchester, Bell,
  D'Amico, McKay \& Crawford}{Camilo et~al.}{2000}]{ckl+00}
Camilo F.,  Kaspi V.~M.,  Lyne A.~G.,  Manchester R.~N.,  Bell J.~F.,  D'Amico
  N.,  McKay N. P.~F.,    Crawford F.,  2000, ApJ, 541, 367

\bibitem[\protect\citeauthoryear{{Camilo}, {Ransom}, {Halpern} \&
  {Reynolds}}{{Camilo} et~al.}{2007}]{crhr07}
{Camilo} F.,  {Ransom} S.~M.,  {Halpern} J.~P.,    {Reynolds} J.,  2007, ApJ,
  666, L93

\bibitem[\protect\citeauthoryear{{Camilo}, {Ransom}, {Halpern}, {Reynolds},
  {Helfand}, {Zimmerman} \& {Sarkissian}}{{Camilo} et~al.}{2006}]{crh+06}
{Camilo} F.,  {Ransom} S.~M.,  {Halpern} J.~P.,  {Reynolds} J.,  {Helfand}
  D.~J.,  {Zimmerman} N.,    {Sarkissian} J.,  2006, Nature, 442, 892

\bibitem[\protect\citeauthoryear{{Camilo}, {Reynolds}, {Johnston}, {Halpern} \&
  {Ransom}}{{Camilo} et~al.}{2008}]{crj+08}
{Camilo} F.,  {Reynolds} J.,  {Johnston} S.,  {Halpern} J.~P.,    {Ransom}
  S.~M.,  2008, ApJ, 679, 681

\bibitem[\protect\citeauthoryear{{Camilo}, {Reynolds}, {Johnston}, {Halpern},
  {Ransom} \& {van Straten}}{{Camilo} et~al.}{2007}]{crj+07}
{Camilo} F.,  {Reynolds} J.,  {Johnston} S.,  {Halpern} J.~P.,  {Ransom} S.~M.,
     {van Straten} W.,  2007, ApJ, 659, L37

\bibitem[\protect\citeauthoryear{{Cordes} \& {Shannon}}{{Cordes} \&
  {Shannon}}{2008}]{cs08}
{Cordes} J.~M.,  {Shannon} R.~M.,  2008, ApJ, 682, 1152

\bibitem[\protect\citeauthoryear{C\'ordova, Hjellming, Mason \&
  Middleditch}{C\'ordova et~al.}{1989}]{chmm89}
C\'ordova F.~A.,  Hjellming R.~M.,  Mason K.~O.,    Middleditch J.,  1989, ApJ,
  345, 451

\bibitem[\protect\citeauthoryear{Crawford, Gaensler, Kaspi, Manchester, Camilo,
  Lyne \& Pivovaroff}{Crawford et~al.}{2001}]{cgk+01}
Crawford F.,  Gaensler B.~M.,  Kaspi V.~M.,  Manchester R.~N.,  Camilo F.,
  Lyne A.~G.,    Pivovaroff M.~J.,  2001, ApJ, 554, 152

\bibitem[\protect\citeauthoryear{{De Luca}, {Caraveo}, {Mereghetti}, {Negroni}
  \& {Bignami}}{{De Luca} et~al.}{2005}]{dcm+05}
{De Luca} A.,  {Caraveo} P.~A.,  {Mereghetti} S.,  {Negroni} M.,    {Bignami}
  G.~F.,  2005, ApJ, 623, 1051

\bibitem[\protect\citeauthoryear{{Deneva}, {etal1} \& {etal2}}{{Deneva}
  et~al.}{2009}]{dcm+09}
{Deneva} J.~S., et al. 2009, ApJ, 703, 2259

\bibitem[\protect\citeauthoryear{{Dib}, {Kaspi} \& {Gavriil}}{{Dib}
  et~al.}{2007}]{dkg07}
{Dib} R.,  {Kaspi} V.~M.,    {Gavriil} F.~P.,  2007, Astrophys. Space Sci.,
  p.~44

\bibitem[\protect\citeauthoryear{{Dib}, {Kaspi} \& {Gavriil}}{{Dib}
  et~al.}{2008}]{dkg08}
{Dib} R.,  {Kaspi} V.~M.,    {Gavriil} F.~P.,  2008, ApJ, 673, 1044

\bibitem[\protect\citeauthoryear{{Edwards}, {Hobbs} \& {Manchester}}{{Edwards}
  et~al.}{2006}]{ehm06}
{Edwards} R.~T.,  {Hobbs} G.~B.,    {Manchester} R.~N.,  2006, MNRAS, 372, 1549

\bibitem[\protect\citeauthoryear{{Edwards} \& {Stappers}}{{Edwards} \&
  {Stappers}}{2002}]{es02}
{Edwards} R.~T.,  {Stappers} B.~W.,  2002, A\&A, 393, 733

\bibitem[\protect\citeauthoryear{{Ershov} \& {Kuzmin}}{{Ershov} \&
  {Kuzmin}}{2005}]{ek05}
{Ershov} A.~A.,  {Kuzmin} A.~D.,  2005, A\&A, 443, 593

\bibitem[\protect\citeauthoryear{{Everett} \& {Weisberg}}{{Everett} \&
  {Weisberg}}{2001}]{ew01}
{Everett} J.~E.,  {Weisberg} J.~M.,  2001, ApJ, 553, 341

\bibitem[\protect\citeauthoryear{{Gaensler}, {McLaughlin}, {Reynolds},
  {Borkowski}, {Rea}, {Possenti}, {Israel}, {Burgay}, {Camilo}, {Chatterjee},
  {Kramer}, {Lyne} \& {Stairs}}{{Gaensler} et~al.}{2007}]{gmr+07}
{Gaensler} B.~M.,  {McLaughlin} M.,  {Reynolds} S.,  {Borkowski} K.,  {Rea} N.,
   {Possenti} A.,  {Israel} G.,  {Burgay} M.,  {Camilo} F.,  {Chatterjee} S.,
  {Kramer} M.,  {Lyne} A.,    {Stairs} I.,  2007, Ap\&SS, 308, 95

\bibitem[\protect\citeauthoryear{{Gavriil}, {Dib} \& {Kaspi}}{{Gavriil}
  et~al.}{2009}]{gdk09}
{Gavriil} F.~P.,  {Dib} R.,    {Kaspi} V.~M.,  2009, ArXiv e-prints (astro-ph/0905.1256)

\bibitem[\protect\citeauthoryear{{Gavriil}, {Gonzalez}, {Gotthelf}, {Kaspi},
  {Livingstone} \& {Woods}}{{Gavriil} et~al.}{2008}]{ggg+08}
{Gavriil} F.~P.,  {Gonzalez} M.~E.,  {Gotthelf} E.~V.,  {Kaspi} V.~M.,
  {Livingstone} M.~A.,    {Woods} P.~M.,  2008, Science, 319, 1802

\bibitem[\protect\citeauthoryear{Gil, Gronkowski \& Rudnicki}{Gil
  et~al.}{1984}]{ggr84}
Gil J.~A.,  Gronkowski P.,    Rudnicki W.,  1984, A\&A, 132, 312

\bibitem[\protect\citeauthoryear{{Gonzalez}, {Kaspi}, {Camilo}, {Gaensler} \&
  {Pivovaroff}}{{Gonzalez} et~al.}{2005}]{gkc+05}
{Gonzalez} M.~E.,  {Kaspi} V.~M.,  {Camilo} F.,  {Gaensler} B.~M.,
  {Pivovaroff} M.~J.,  2005, ApJ, 630, 489

\bibitem[\protect\citeauthoryear{{Hankins}, {Kern}, {Weatherall} \&
  {Eilek}}{{Hankins} et~al.}{2003}]{hkw+03}
{Hankins} T.~H.,  {Kern} J.~S.,  {Weatherall} J.~C.,    {Eilek} J.~A.,  2003,
  Nature, 422

\bibitem[\protect\citeauthoryear{{Hessels}, {Ransom}, {Kaspi}, {Roberts},
  {Champion} \& {Stappers}}{{Hessels} et~al.}{2008}]{hrk+08}
{Hessels} J.~W.~T.,  {Ransom} S.~M.,  {Kaspi} V.~M.,  {Roberts} M.~S.~E.,
  {Champion} D.~J.,    {Stappers} B.~W.,  2008, in {C.~Bassa, Z.~Wang,
  A.~Cumming, \& V.~M.~Kaspi} ed., 40 Years of Pulsars: Millisecond Pulsars,
  Magnetars and More Vol.~983 of American Institute of Physics Conference
  Series, {The GBT350 Survey of the Northern Galactic Plane for Radio Pulsars
  and Transients}.
pp 613--615

\bibitem[\protect\citeauthoryear{{Hobbs}, {Edwards} \& {Manchester}}{{Hobbs}
  et~al.}{2006}]{hem06}
{Hobbs} G.~B.,  {Edwards} R.~T.,    {Manchester} R.~N.,  2006, MNRAS, 369, 655

\bibitem[\protect\citeauthoryear{{Johnston}, {Karastergiou} \&
  {Willett}}{{Johnston} et~al.}{2006}]{jkw06}
{Johnston} S.,  {Karastergiou} A.,    {Willett} K.,  2006, MNRAS, 369, 1916

\bibitem[\protect\citeauthoryear{Johnston \& Romani}{Johnston \&
  Romani}{2002}]{jr02}
Johnston S.,  Romani R.,  2002, MNRAS, 332, 109

\bibitem[\protect\citeauthoryear{{Johnston}, {van Straten}, {Kramer} \&
  {Bailes}}{{Johnston} et~al.}{2001}]{jvkb01}
{Johnston} S.,  {van Straten} W.,  {Kramer} M.,    {Bailes} M.,  2001, ApJ,
  549, L101

\bibitem[\protect\citeauthoryear{{Johnston} \& {Weisberg}}{{Johnston} \&
  {Weisberg}}{2006}]{jw06}
{Johnston} S.,  {Weisberg} J.~M.,  2006, MNRAS, 368, 1856

\bibitem[\protect\citeauthoryear{{Kaplan}}{{Kaplan}}{2008}]{kap08}
{Kaplan} D.~L.,  2008, in {Y.-F.~Yuan, X.-D.~Li, \& D.~Lai} ed., Astrophysics
  of Compact Objects Vol.~968 of American Institute of Physics Conference
  Series, {Nearby, Thermally Emitting Neutron Stars}.
pp 129--136

\bibitem[\protect\citeauthoryear{{Karastergiou}, {Hotan}, {van Straten},
  {McLaughlin} \& {Ord}}{{Karastergiou} et~al.}{2009}]{khs+09}
{Karastergiou} A.,  {Hotan} A.~W.,  {van Straten} W.,  {McLaughlin} M.~A.,
  {Ord} S.~M.,  2009, MNRAS, 396, L95

\bibitem[\protect\citeauthoryear{{Keane} \& {Kramer}}{{Keane} \&
  {Kramer}}{2008}]{kk08}
{Keane} E.~F.,  {Kramer} M.,  2008, MNRAS, 391, 2009

\bibitem[\protect\citeauthoryear{{Keane}, {Ludovici}, {Eatough}, {Kramer},
  {Lyne}, {McLaughlin} \& {Stappers}}{{Keane} et~al.}{2010}]{kle+10}
{Keane} E.~F.,  {Ludovici} D.~A.,  {Eatough} R.~P.,  {Kramer} M.,  {Lyne}
  A.~G.,  {McLaughlin} M.~A.,    {Stappers} B.~W.,  2010, MNRAS, 401, 1057

\bibitem[\protect\citeauthoryear{{Kramer}, {Karastergiou}, {Gupta}, {Johnston},
  {Bhat} \& {Lyne}}{{Kramer} et~al.}{2003}]{kkg+03}
{Kramer} M.,  {Karastergiou} A.,  {Gupta} Y.,  {Johnston} S.,  {Bhat} N.~D.~R.,
     {Lyne} A.~G.,  2003, A\&A, 407, 655

\bibitem[\protect\citeauthoryear{{Kramer}, {Lyne}, {O'Brien}, {Jordan} \&
  {Lorimer}}{{Kramer} et~al.}{2006}]{klo+06}
{Kramer} M.,  {Lyne} A.~G.,  {O'Brien} J.~T.,  {Jordan} C.~A.,    {Lorimer}
  D.~R.,  2006, Science, 312, 549

\bibitem[\protect\citeauthoryear{{Kramer}, {Stappers}, {Jessner}, {Lyne} \&
  {Jordan}}{{Kramer} et~al.}{2007}]{ksj+07}
{Kramer} M.,  {Stappers} B.~W.,  {Jessner} A.,  {Lyne} A.~G.,    {Jordan}
  C.~A.,  2007, MNRAS, 377, 107

\bibitem[\protect\citeauthoryear{{Kuiper} \& {Hermsen}}{{Kuiper} \&
  {Hermsen}}{2009}]{kh09}
{Kuiper} L.,  {Hermsen} W.,  2009, A\&A, 501, 1031

\bibitem[\protect\citeauthoryear{{Kumar} \& {Safi-Harb}}{{Kumar} \&
  {Safi-Harb}}{2008}]{ks08}
{Kumar} H.~S.,  {Safi-Harb} S.,  2008, ApJ, 678, L43

\bibitem[\protect\citeauthoryear{{Levin}, {Bailes}, {Bates}, {Bhat}, {Burgay},
  {Burke-Spolaor}, {D'Amico}, {Johnston}, {Keith}, {Kramer}, {Milia},
  {Possenti}, {Rea}, {Stappers} \& {van Straten}}{{Levin}
  et~al.}{2010}]{lbb+10}
{Levin} L.,  {Bailes} M.,  {Bates} S.,  {Bhat} N.~D.~R.,  {Burgay} M.,
  {Burke-Spolaor} S.,  {D'Amico} N.,  {Johnston} S.,  {Keith} M.,  {Kramer} M.,
   {Milia} S.,  {Possenti} A.,  {Rea} N.,  {Stappers} B.,    {van Straten} W.,
  2010, ApJ, 721, L33

\bibitem[\protect\citeauthoryear{{Livingstone}, {Kaspi} \&
  {Gavriil}}{{Livingstone} et~al.}{2010}]{lkg10}
{Livingstone} M.~A.,  {Kaspi} V.~M.,    {Gavriil} F.~P.,  2010, ApJ, 710, 1710

\bibitem[\protect\citeauthoryear{{Livingstone}, {Kaspi}, {Gavriil},
  {Manchester}, {Gotthelf} \& {Kuiper}}{{Livingstone} et~al.}{2007}]{lkg+07}
{Livingstone} M.~A.,  {Kaspi} V.~M.,  {Gavriil} F.~P.,  {Manchester} R.~N.,
  {Gotthelf} E.~V.~G.,    {Kuiper} L.,  2007, Astrophys. Space Sci., 308, 317

\bibitem[\protect\citeauthoryear{{Livingstone}, {Kaspi}, {Gotthelf} \&
  {Kuiper}}{{Livingstone} et~al.}{2006}]{lkgk06}
{Livingstone} M.~A.,  {Kaspi} V.~M.,  {Gotthelf} E.~V.,    {Kuiper} L.,  2006,
  ApJ, 647, 1286

\bibitem[\protect\citeauthoryear{{Lorimer}, {Bailes}, {McLaughlin}, {Narkevic}
  \& {Crawford}}{{Lorimer} et~al.}{2007}]{lbm+07}
{Lorimer} D.~R.,  {Bailes} M.,  {McLaughlin} M.~A.,  {Narkevic} D.~J.,
  {Crawford} F.,  2007, Science, 318, 777

\bibitem[\protect\citeauthoryear{{Lorimer} \& {Kramer}}{{Lorimer} \&
  {Kramer}}{2005}]{lk05}
{Lorimer} D.~R.,  {Kramer} M.,  2005, {Handbook of Pulsar Astronomy}.
Cambridge University Press

\bibitem[\protect\citeauthoryear{Lundgren, Cordes, Ulmer, Matz, Lomatch, Foster
  \& Hankins}{Lundgren et~al.}{1995}]{lcu+95}
Lundgren S.~C.,  Cordes J.~M.,  Ulmer M.,  Matz S.~M.,  Lomatch S.,  Foster
  R.~S.,    Hankins T.,  1995, ApJ, 453, 433

\bibitem[\protect\citeauthoryear{{Luo} \& {Melrose}}{{Luo} \&
  {Melrose}}{2007}]{lm07}
{Luo} Q.,  {Melrose} D.,  2007, MNRAS, 378, 1481

\bibitem[\protect\citeauthoryear{{Lyne}, {Hobbs}, {Kramer}, {Stairs} \&
  {Stappers}}{{Lyne} et~al.}{2010}]{lhk+10}
{Lyne} A.,  {Hobbs} G.,  {Kramer} M.,  {Stairs} I.,    {Stappers} B.,  2010,
  Science, 329, 408

\bibitem[\protect\citeauthoryear{{Lyne}}{{Lyne}}{2004}]{lyn04}
{Lyne} A.~G.,  2004, in {F.~Camilo \& B.~M.~Gaensler} ed., Young Neutron Stars
  and Their Environments Vol.~218 of IAU Symposium, {From Crab Pulsar to
  Magnetar?}.
pp 257--+

\bibitem[\protect\citeauthoryear{{Lyne}, {McLaughlin}, {Keane}, {Kramer},
  {Espinoza}, {Stappers}, {Palliyaguru} \& {Miller}}{{Lyne}
  et~al.}{2009}]{lmk+09}
{Lyne} A.~G.,  {McLaughlin} M.~A.,  {Keane} E.~F.,  {Kramer} M.,  {Espinoza}
  C.~M.,  {Stappers} B.~W.,  {Palliyaguru} N.~T.,    {Miller} J.,  2009, MNRAS,
  400, 1439

\bibitem[\protect\citeauthoryear{{Lyne}, {Shemar} \& {Graham-Smith}}{{Lyne}
  et~al.}{2000}]{lsg00}
{Lyne} A.~G.,  {Shemar} S.~L.,    {Graham-Smith} F.,  2000, MNRAS, 315, 534

\bibitem[\protect\citeauthoryear{McCulloch, Hamilton, McConnell \&
  King}{McCulloch et~al.}{1990}]{mhmk90}
McCulloch P.~M.,  Hamilton P.~A.,  McConnell D.,    King E.~A.,  1990, Nature,
  346, 822

\bibitem[\protect\citeauthoryear{McKenna \& Lyne}{McKenna \& Lyne}{1990}]{ml90}
McKenna J.,  Lyne A.~G.,  1990, Nature, 343, 349

\bibitem[\protect\citeauthoryear{{Manchester}, {etal1} \& {etal2}}{{Manchester}
  et~al.}{2001}]{mlc+01}
{Manchester} R.~N., et al. 2001, MNRAS, 328, 17

\bibitem[\protect\citeauthoryear{{McLaughlin} \& {Cordes}}{{McLaughlin} \&
  {Cordes}}{2003}]{mc03}
{McLaughlin} M.~A.,  {Cordes} J.~M.,  2003, ApJ, 596, 982

\bibitem[\protect\citeauthoryear{{McLaughlin}, {etal1} \& {etal2}}{{McLaughlin}
  et~al.}{2006}]{mll+06}
{McLaughlin} M.~A., et al. 2006, Nature, 439, 817

\bibitem[\protect\citeauthoryear{{McLaughlin}, {Lyne}, {Keane}, {Kramer},
  {Miller}, {Lorimer}, {Manchester}, {Camilo} \& {Stairs}}{{McLaughlin}
  et~al.}{2009}]{mlk+09}
{McLaughlin} M.~A.,  {Lyne} A.~G.,  {Keane} E.~F.,  {Kramer} M.,  {Miller}
  J.~J.,  {Lorimer} D.~R.,  {Manchester} R.~N.,  {Camilo} F.,    {Stairs}
  I.~H.,  2009, MNRAS, 400, 1431

\bibitem[\protect\citeauthoryear{{McLaughlin}, {Rea}, {Gaensler}, {Chatterjee},
  {Camilo}, {Kramer}, {Lorimer}, {Lyne}, {Israel} \& {Possenti}}{{McLaughlin}
  et~al.}{2007}]{mrg+07}
{McLaughlin} M.~A.,  {Rea} N.,  {Gaensler} B.~M.,  {Chatterjee} S.,  {Camilo}
  F.,  {Kramer} M.,  {Lorimer} D.~R.,  {Lyne} A.~G.,  {Israel} G.~L.,
  {Possenti} A.,  2007, ApJ, 670, 1307

\bibitem[\protect\citeauthoryear{{Melatos} \& {Warszawski}}{{Melatos} \&
  {Warszawski}}{2009}]{mw09}
{Melatos} A.,  {Warszawski} L.,  2009, ApJ, 700, 1524

\bibitem[\protect\citeauthoryear{{Press}, {Teukolsky}, {Vetterling} \&
  {Flannery}}{{Press} et~al.}{1992}]{ptv+92}
{Press} W.~H.,  {Teukolsky} S.~A.,  {Vetterling} W.~T.,    {Flannery} B.~P.,
  1992, {Numerical recipes in C. The art of scientific computing}.
Cambridge: University Press, |c1992, 2nd ed.

\bibitem[\protect\citeauthoryear{Radhakrishnan \& Cooke}{Radhakrishnan \&
  Cooke}{1969}]{rc69a}
Radhakrishnan V.,  Cooke D.~J.,  1969, Astrophys. Lett., 3, 225

\bibitem[\protect\citeauthoryear{Radhakrishnan \& Manchester}{Radhakrishnan \&
  Manchester}{1969}]{rm69}
Radhakrishnan V.,  Manchester R.~N.,  1969, Nature, 222, 228

\bibitem[\protect\citeauthoryear{Rankin}{Rankin}{1986}]{ran86}
Rankin J.~M.,  1986, ApJ, 301, 901

\bibitem[\protect\citeauthoryear{{Reynolds}, {Borkowski}, {Gaensler}, {Rea},
  {McLaughlin}, {Possenti}, {Israel}, {Burgay}, {Camilo}, {Chatterjee},
  {Kramer}, {Lyne} \& {Stairs}}{{Reynolds} et~al.}{2006}]{rbg+06}
{Reynolds} S.~P.,  {Borkowski} K.~J.,  {Gaensler} B.~M.,  {Rea} N.,
  {McLaughlin} M.,  {Possenti} A.,  {Israel} G.,  {Burgay} M.,  {Camilo} F.,
  {Chatterjee} S.,  {Kramer} M.,  {Lyne} A.,    {Stairs} I.,  2006, ApJ, 639,
  L71

\bibitem[\protect\citeauthoryear{Ruderman}{Ruderman}{1969}]{rud69}
Ruderman M.,  1969, Nature, 223, 597

\bibitem[\protect\citeauthoryear{Ruderman, Zhu \& Chen}{Ruderman
  et~al.}{1998}]{rzc98}
Ruderman M.,  Zhu T.,    Chen K.,  1998, ApJ, 492, 267

\bibitem[\protect\citeauthoryear{{Serylak}, {Stappers}, {Weltevrede}, {Kramer},
  {Jessner}, {Lyne}, {Jordan}, {Lazaridis} \& {Zensus}}{{Serylak}
  et~al.}{2009}]{ssw+09}
{Serylak} M.,  {Stappers} B.~W.,  {Weltevrede} P.,  {Kramer} M.,  {Jessner} A.,
   {Lyne} A.~G.,  {Jordan} C.~A.,  {Lazaridis} K.,    {Zensus} J.~A.,  2009,
  MNRAS, 394, 295

\bibitem[\protect\citeauthoryear{Shemar \& Lyne}{Shemar \& Lyne}{1996}]{sl96}
Shemar S.~L.,  Lyne A.~G.,  1996, MNRAS, 282, 677

\bibitem[\protect\citeauthoryear{{Smith}, {etal1} \& {etal2}}{{Smith}
  et~al.}{2008}]{sgc+08}
{Smith} D.~A., et al. 2008, A\&A, 492, 923

\bibitem[\protect\citeauthoryear{{Standish}}{{Standish}}{1998}]{sta98c}
{Standish} E.~M.,  1998, A\&A, 336, 381

\bibitem[\protect\citeauthoryear{Taylor, Manchester \& Huguenin}{Taylor
  et~al.}{1975}]{tmh75}
Taylor J.~H.,  Manchester R.~N.,    Huguenin G.~R.,  1975, ApJ, 195, 513

\bibitem[\protect\citeauthoryear{von Hoensbroech, Lesch \& Kunzl}{von
  Hoensbroech et~al.}{1998}]{hlk98}
von Hoensbroech A.,  Lesch H.,    Kunzl T.,  1998, A\&A, 336, 209

\bibitem[\protect\citeauthoryear{{Wang}, {Manchester} \& {Johnston}}{{Wang}
  et~al.}{2007}]{wmj07}
{Wang} N.,  {Manchester} R.~N.,    {Johnston} S.,  2007, MNRAS, 377, 1383

\bibitem[\protect\citeauthoryear{Wang, Manchester, Pace, Bailes, Kaspi,
  Stappers \& Lyne}{Wang et~al.}{2000}]{wmp+00}
Wang N.,  Manchester R.~N.,  Pace R.,  Bailes M.,  Kaspi V.~M.,  Stappers
  B.~W.,    Lyne A.~G.,  2000, MNRAS, 317, 843

\bibitem[\protect\citeauthoryear{{Weltevrede}, {Edwards} \&
  {Stappers}}{{Weltevrede} et~al.}{2006}]{wes06}
{Weltevrede} P.,  {Edwards} R.~T.,    {Stappers} B.~W.,  2006, A\&A, 445, 243

\bibitem[\protect\citeauthoryear{{Weltevrede}, {etal1} \& {etal2}}{{Weltevrede}
  et~al.}{2010}]{waa+10}
{Weltevrede} P., et al. 2010, ApJ, 708, 1426

\bibitem[\protect\citeauthoryear{{Weltevrede} \& {Johnston}}{{Weltevrede} \&
  {Johnston}}{2008}]{wj08b}
{Weltevrede} P.,  {Johnston} S.,  2008, MNRAS, 391, 1210

\bibitem[\protect\citeauthoryear{{Weltevrede}, {Johnston}, {Manchester},
  {Bhat}, {Burgay}, {Champion}, {Hobbs}, {K{\i}z{\i}ltan}, {Keith}, {Possenti},
  {Reynolds} \& {Watters}}{{Weltevrede} et~al.}{2010}]{wjm+10}
{Weltevrede} P.,  {Johnston} S.,  {Manchester} R.~N.,  {Bhat} R.,  {Burgay} M.,
   {Champion} D.,  {Hobbs} G.~B.,  {K{\i}z{\i}ltan} B.,  {Keith} M.,
  {Possenti} A.,  {Reynolds} J.~E.,    {Watters} K.,  2010, PASA, 27, 64

\bibitem[\protect\citeauthoryear{{Weltevrede}, {Stappers}, {Rankin} \&
  {Wright}}{{Weltevrede} et~al.}{2006}]{wsr+06}
{Weltevrede} P.,  {Stappers} B.~W.,  {Rankin} J.~M.,    {Wright} G.~A.~E.,
  2006, ApJ, 645, L149

\bibitem[\protect\citeauthoryear{{Weltevrede}, {Wright}, {Stappers} \&
  {Rankin}}{{Weltevrede} et~al.}{2006}]{wws+06}
{Weltevrede} P.,  {Wright} G.~A.~E.,  {Stappers} B.~W.,    {Rankin} J.~M.,
  2006, A\&A, 458, 269

\bibitem[\protect\citeauthoryear{Wong, Backer \& Lyne}{Wong
  et~al.}{2001}]{wbl01}
Wong T.,  Backer D.~C.,    Lyne A.,  2001, ApJ, 548, 447

\bibitem[\protect\citeauthoryear{{Woods} \& {Thompson}}{{Woods} \&
  {Thompson}}{2006}]{wt06}
{Woods} P.~M.,  {Thompson} C.,  2006, Compact Stellar X-ray Sources.
Cambridge University Press

\bibitem[\protect\citeauthoryear{{Yuan}, {Manchester}, {Wang}, {Zhou}, {Liu} \&
  {Gao}}{{Yuan} et~al.}{2010}]{ymw+10}
{Yuan} J.~P.,  {Manchester} R.~N.,  {Wang} N.,  {Zhou} X.,  {Liu} Z.~Y.,
  {Gao} Z.~F.,  2010, ApJ, 719, L111

\end{thebibliography}

\label{lastpage} 

\clearpage

\end{document}